\documentclass[aps,prb,twocolumn,superscriptaddress,longbibliography,nofootinbib]{revtex4-2}

\usepackage[utf8]{inputenc}
\usepackage[T1]{fontenc}
\usepackage{amsmath,amssymb,amsfonts,bm}
\usepackage{graphicx}
\usepackage{xcolor}
\usepackage{hyperref}
\usepackage{mathtools}

\hypersetup{colorlinks=true,linkcolor=blue,citecolor=blue,urlcolor=blue}

\DeclareMathOperator{\sgn}{sgn}

\newcommand{\I}{{\rm i}}

\renewcommand{\Re}{{\rm Re}}
\renewcommand{\Im}{{\rm Im}}

\newcommand{\dd}{\mathrm{d}}

\newcommand{\bx}{{\bf x}}

\newcommand{\hbz}{\hat{\bf z}}

\newcommand{\bq}{{\bf q}}
\newcommand{\bk}{{\bf k}}
\newcommand{\bd}{{\bf d}}

\newcommand{\bE}{{\bf E}}

\newcommand{\bQ}{{\bf Q}}
\newcommand{\brho}{\mbox{\boldmath $x_\rho$}}

\newcommand{\balpha}{\mbox{\boldmath $\alpha$}}

\newcommand{\calD}{\mathcal{D}}
\newcommand{\calG}{\mathcal{G}}
\newcommand{\calR}{\mathcal{R}}
\newcommand{\obG}{\overleftrightarrow{\bf G}}
\newcommand{\obF}{\overleftrightarrow{\bf F}}
\newcommand{\bI}{{\bf I}}

\newcommand{\TM}{\mathrm{TM}}
\newcommand{\TE}{\mathrm{TE}}

\newcommand{\be}{\begin{equation}}
\newcommand{\ee}{\end{equation}}
\newcommand{\nn}{\nonumber}

\begin{document}

\title{Universal tuning of F\"{o}rster resonance energy transfer in gate-programmable conductor-dielectric-conductor heterostructures}

\author{Alexis J. Agosto}
\affiliation{NanoScience Technology Center, Department of Physics, University of Central Florida, Orlando, Florida 32826, USA}
\author{Daniel Gunlycke}
\affiliation{U.S. Naval Research Laboratory, 4555 Overlook Ave SW, Washington, DC 20375, USA}
\email{lennart.d.gunlycke.civ@us.navy.mil}
\author{Michael N. Leuenberger}
\affiliation{NanoScience Technology Center, Department of Physics, University of Central Florida, Orlando, Florida 32826, USA}
\affiliation{College of Optics and Photonics, University of Central Florida, Orlando, Florida 32826, USA}
\email{michael.leuenberger@ucf.edu}
\date{\today}

\begin{abstract}
We develop a quantum-electrodynamical theory for universal tuning of spontaneous emission and F\"{o}rster resonance energy transfer (FRET) in a material-agnostic conductor-dielectric-conductor heterostructure.  The platform consists of a dielectric spacer of thickness $W$ bounded by two gate-tunable two-dimensional conductors.  The only microscopic input from the surrounding materials is the transverse-magnetic and transverse-electric reflection amplitudes $r_{\TM/\TE}(q_\rho,\omega)$ of the two sheets.  Starting from the QED photon propagator, we derive the retarded Maxwell dyadic, the vacuum/Hadamard field propagator, and the time-ordered Feynman propagator in the same geometry.  The spontaneous-emission rate is controlled by the local vacuum spectral density, equivalently by $\Im\,\obG_{\text R}(\bx_0,\bx_0;\omega_0)$, while FRET is controlled by the nonlocal retarded/advanced product $\obG_{\text R}(\bx_D,\bx_A;\omega_D)\Im\,\balpha_A(\omega_D)\obG_{\text A}(\bx_A,\bx_D;\omega_D)$.  Multiple reflections between the two conductors are resummed exactly in a reduced one-dimensional problem along $z$.  In the transparent limit $r_{\TM}\to 0$, the near-field FRET rate recovers the bulk $x_\rho^{-6}$ law.  In the Dirichlet/PEC branch $r_{\TM}\to -1$, the gapless transverse mode is removed and the donor-acceptor coupling acquires a Bessel-$K$ envelope, giving an exponentially screened FRET rate $\Gamma_{D\to A}\propto\exp(-2\pi x_\rho/W)$ at large lateral separation.  In the opposite Neumann/PMC-like branch $r_{\TM}\to 1^{-}$, a nearly gapless transverse mode survives and produces a wide quasi-two-dimensional logarithmic propagator, enhancing the nonlocal electromagnetic coupling over a gate-programmable range $x_{\rho,*}\sim W/(1-r_{\TM})$.  For graphene-Er implementations, the same retarded Green tensor also separates dissipative on-shell Er-to-graphene decay, governed by its absorptive part, from dispersive virtual-plasmon-mediated Er-Er coupling, governed by its reactive part; a plasmonic band gap can suppress the former while retaining the latter.  The resulting formalism provides a route to electrical control of radiative decay, near-field energy transfer, and exciton-exciton couplings in moir\'e and van der Waals quantum materials.
\end{abstract}

\maketitle

\section{Introduction}

F\"{o}rster resonance energy transfer (FRET) is the near-field migration of an electronic excitation from an excited donor to an acceptor through electromagnetic dipole-dipole coupling.  In homogeneous three-dimensional space the original F\"{o}rster theory gives the familiar $R^{-6}$ transfer rate for electric-dipole transitions, while Dexter transfer describes the short-range exchange-mediated channel \cite{Forster1948,Dexter1953}.  The electromagnetic environment can strongly reshape both the donor decay rate and the donor-acceptor transfer rate.  This has long been appreciated for emitters near interfaces and multilayers \cite{ChanceProckSilbey1978,WylieSipe1984,WylieSipe1985}, and it has a compact formulation in macroscopic quantum electrodynamics (MQED), where both spontaneous emission and resonance energy transfer are expressed in terms of the dyadic Green tensor of Maxwell's equations \cite{DungKnollWelsch2002,ScheelBuhmann2008,Buhmann2012I,NovotnyHecht2012,JonesBradshaw2019}.

A particularly important experimental benchmark is the work of Cano \textit{et al.}, who demonstrated fast electrical control of the near-field interaction between Er$^{3+}$ emitters and graphene at the telecom transition near $1.54~\mu{\rm m}$ \cite{Cano2020ErGraphene}.  They observed decay-rate enhancements exceeding $10^3$ for approximately $25\%$ of the emitters and modulated the Er-graphene interaction electrically at frequencies up to $300~{\rm kHz}$.  Their measurements also resolved two distinct dissipative Er-to-graphene channels: for $E_F\lesssim0.4~{\rm eV}$, interband electron-hole-pair excitation dominates, whereas above the Pauli-blocking threshold the intraband response becomes dominant and exhibits a clear graphene-plasmon contribution.  These results establish that the Er-graphene near-field interaction can be both exceptionally strong and rapidly gate tunable.

Here we propose to use the conductor-dielectric-conductor architecture introduced in our earlier work on universal QED interaction tuning \cite{LeuenbergerGunlycke2026} as a programmable platform for FRET.  Two parallel, gate-tunable, two-dimensional conductors bound a dielectric spacer.  Their frequency- and wave-vector-dependent reflection amplitudes $r_{\TM/\TE}(q_\rho,\omega)$ reshape the electromagnetic propagator.  Because all radiative and near-field processes can be written in terms of this propagator, a single material-agnostic geometry can tune spontaneous emission and FRET.  Importantly, high-conductivity multilayer graphene mirrors have several concrete materials realizations, including rotationally decoupled C-face epitaxial graphene on SiC, FeCl$_3$-intercalated few-layer graphene, and more recently $\alpha$-RuCl$_3$-intercalated graphite \cite{deHeer2007EpitaxialGraphene,Hass2008MultilayerGraphene,Orlita2008HighMobilityMEG,Zhan2010FeCl3FLG,Khrapach2012FeCl3Conductors,Razpopov2026RuCl3Intercalated}.

The experiment of Cano \textit{et al.} also motivates a distinction that is central to the present work \cite{Cano2020ErGraphene}.  When graphene provides on-shell electronic or plasmonic states at the Er transition frequency, the Er-graphene interaction is dissipative: a real electron-hole excitation or propagating graphene plasmon is created, and the corresponding decay is encoded in the absorptive part of the retarded Green tensor.  By contrast, when the relevant plasmon pole is sufficiently detuned or removed from the Er frequency by a plasmonic band gap, the same microscopic light-matter coupling acts predominantly dispersively through the reactive part of the Green tensor.  In that regime graphene plasmons can mediate virtual Er-Er coupling while irreversible Er-to-graphene energy flow is strongly reduced.  Our goal is therefore not simply to maximize the Er-graphene interaction, but to engineer the relative weights of its dissipative and dispersive components.

Here we show that spontaneous emission is determined by the vacuum propagator, which gives the field correlation that drives spontaneous emission, and that FRET is determined by
the retarded Green function, which gives the causal donor-to-acceptor electromagnetic response that enters FRET.
The time-ordered Feynman propagator remains the natural object in perturbation theory, but physical rates reorganize into retarded/advanced and absorptive response functions after the intermediate photon and acceptor states are summed.

The paper is organized as follows.  Section~\ref{sec:geometry} defines the geometry and reflection-amplitude input.  Section~\ref{sec:kernels} derives the relation between the Feynman, retarded, advanced, and vacuum electric-field propagators.  Section~\ref{sec:cavitygreen} gives the exact multiple-reflection reduced Green function and its Poisson-resummed limiting forms.  Section~\ref{sec:spontaneous} derives the spontaneous-emission rate from the vacuum propagator.  Section~\ref{sec:fret} derives the FRET rate from the retarded Green tensor.  Section~\ref{sec:tuning} extracts the universal tuning laws: bulk power-law transfer, exponential screening, and logarithmic antiscreening.  Detailed derivations are organized in the Supplementary Information as follows: Sec.~S1 gives the QED action and photon-propagator conventions; Sec.~S2 derives the retarded cylindrical/Weyl reduction and pole structure; Sec.~S3 develops the surface-mode hybridization and self-energy formulation; Sec.~S4 derives the TM/TE boundary conditions and graphene reflection amplitudes; Sec.~S5 gives the vacuum-correlation derivation of spontaneous emission; Sec.~S6 derives the FRET transition amplitude, rate, and spectral-overlap form; Secs.~S7 and S8 develop the signed image lattice, dyadic derivatives, Poisson/Bessel resummation, and universal tuning laws; Secs.~S9 and S10 give the analytical Er--Er formulas and numerical workflow; Sec.~S11 gives the semi-empirical donor--acceptor spectral model; and Sec.~S12 surveys material platforms for tuning the signed TM reflection amplitude.  Complementary derivations retained in the main manuscript are given in Appendix~\ref{app:Tmatrix} (transition operator), Appendix~\ref{app:contours} (contours and cylindrical waves), Appendix~\ref{app:poisson} (Poisson summation), Appendix~\ref{app:dyadic} (dyadic derivatives), Appendix~\ref{app:finiteT} (finite temperature and MQED), Appendix~\ref{app:numerics} (numerical design rules), and Appendix~\ref{app:metagate_tuned_DLG_green_functions} (metagate-tuned double-layer-graphene Green functions).

\begin{figure}[htb]
\centering
\includegraphics[width=0.50\textwidth]{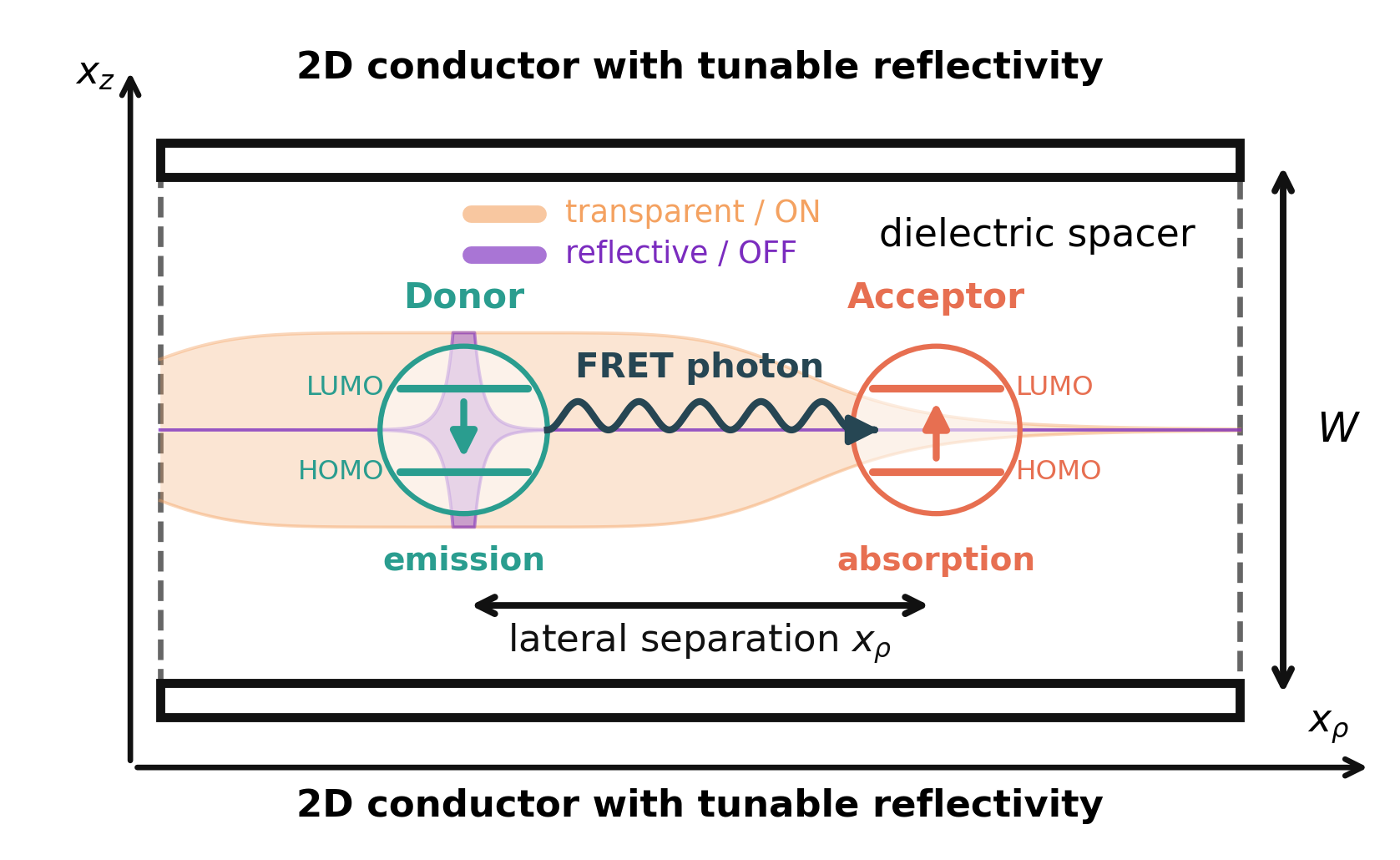}
\caption{
Schematic of tunable F\"orster resonance energy transfer (FRET) in a
conductor-dielectric-conductor heterostructure.  A dielectric spacer of
thickness \(W\) is bounded by two gate-tunable two-dimensional conductors,
whose electromagnetic response is encoded in the reflection amplitudes
\(r_{\rm TM/TE}(q_\rho,\omega)\).  A donor and an acceptor are embedded in
the spacer and separated laterally by \(x_\rho\).  The donor undergoes an
emission transition from its LUMO to its HOMO, while the acceptor undergoes
the corresponding absorption transition from its HOMO to its LUMO.  The
wavy arrow denotes the retarded photon-mediated FRET channel, governed by
the nonlocal Green tensor
\(\obG_{\rm R}(\bx_A,\bx_D;\omega_D)\).  The broad envelope illustrates the
transparent or plasmon-assisted regime, in which the donor field remains
long-ranged, whereas the narrow envelope illustrates the reflective
screening regime, in which the conductor response drives the relevant TM
reflection branch toward the Dirichlet-like limit and exponentially
suppresses long-distance FRET.  Thus, changing the reflectivity of the
two-dimensional conductors provides an in situ knob for tuning the range
and magnitude of donor-acceptor energy transfer without changing the
emitters themselves.
}
\label{fig:FRET_schematic_tunable_range}
\end{figure}

Figure~\ref{fig:FRET_schematic_tunable_range} summarizes the geometry and the central control principle: gate tuning changes the signed reflection amplitudes of the two conducting boundaries and thereby reprograms the range and magnitude of the donor--acceptor electromagnetic coupling.

\section{Geometry and material-agnostic optical input}
\label{sec:geometry}

We consider a dielectric spacer occupying
\begin{equation}
 -\frac{W}{2}<x_z<\frac{W}{2},
\end{equation}
with relative permittivity \(\varepsilon_c(\omega)\) and refractive index
\(n_c(\omega)=\sqrt{\varepsilon_c(\omega)}\).  The spacer is bounded by two
parallel, atomically thin conducting sheets located at \(x_z=\pm W/2\).  A
donor \(D\) and an acceptor \(A\) are located at
\begin{equation}
 \bx_D=(\brho_D,x_{Dz}),\qquad
 \bx_A=(\brho_A,x_{Az}),
\end{equation}
and their in-plane separation is
\begin{equation}
 x_\rho\equiv|\brho_A-\brho_D|.
\end{equation}
Most analytical results below are written for the midplane configuration
\begin{equation}
 x_{Dz}=x_{Az}=0,
\end{equation}
which is the geometry that most clearly exposes the parity-selected
transverse harmonics.  The off-midplane case follows from the same reduced
Green function before setting \(x_z=x_z'=0\).

Because the structure is translationally invariant in the in-plane
directions, each electromagnetic field component can be decomposed into
in-plane momentum \(\bq_\rho\) and polarization
\(\lambda\in\{\TM,\TE\}\).  We write
\be
 q_\rho\equiv |\bq_\rho|,
 \qquad
k_c(\omega)=\frac{n_c(\omega)\omega}{c},
\ee
and
\be
 Q_c(q_\rho,\omega)= \left[\varepsilon_c\left(\frac{\omega}{c}+\I0^+\right)^2-q_\rho^2\right]^{1/2},
\label{eq:Qc_real_def}
\ee
with the retarded branch chosen such that \(\Im Q_c\ge0\).  For evanescent
components we write
\begin{equation}
 \bar Q_c(q_\rho,\omega)=
 \sqrt{q_\rho^2-\varepsilon_c(\omega/c)^2}>0 ,
\label{eq:Qbar_def}
\end{equation}
where $Q_c(q_\rho,\omega)=\I\bar Q_c(q_\rho,\omega)$.
At imaginary frequency \(\omega=\I\xi\), the same decay constant becomes
\begin{equation}
 \bar Q_c(q_\rho,\xi)=
 \sqrt{q_\rho^2+\varepsilon_c(\I\xi)\frac{\xi^2}{c^2}}.
\label{eq:Qbar_imag_def}
\end{equation}

The microscopic optical response of the two sheets enters only through
reflection amplitudes
\begin{equation}
 r_{\lambda}^{\pm}(q_\rho,\omega),\qquad \lambda=\TM,\TE,
\end{equation}
for waves inside the spacer incident on the upper \((+)\) or lower \((-)\)
sheet.  For identical sheets we write
\(r_{\lambda}^{+}=r_{\lambda}^{-}\equiv r_{\lambda}\).  A commonly used
local sheet-conductivity form is
\begin{align}
 r_{\TM}(q_\rho,\omega)
 &=
 \frac{\varepsilon_2 Q_1-\varepsilon_1 Q_2
      -\dfrac{\mu_0 c^2 q_\rho^2}{\omega}\sigma(q_\rho,\omega)}
      {\varepsilon_2 Q_1+\varepsilon_1 Q_2
      +\dfrac{\mu_0 c^2 q_\rho^2}{\omega}\sigma(q_\rho,\omega)},
\label{eq:rTM_sigma_standard}\\
 r_{\TE}(q_\rho,\omega)
 &=
 \frac{Q_1-Q_2-\mu_0\omega\sigma(q_\rho,\omega)}
      {Q_1+Q_2+\mu_0\omega\sigma(q_\rho,\omega)},
\label{eq:rTE_sigma_standard}
\end{align}
where \(Q_j=[\varepsilon_j(\omega/c+\I0^+)^2-q_\rho^2]^{1/2}\) is the
retarded longitudinal wave number in medium \(j\).  Equivalent forms can be
written in terms of the longitudinal density response
\(\chi_{2\mathrm D}(q_\rho,\omega)\) using charge conservation,
\begin{equation}
 \sigma(q_\rho,\omega)=
 -\frac{\I\omega}{q_\rho^2}\chi_{2\mathrm D}(q_\rho,\omega).
\end{equation}
In the electrostatic limit \(\omega\to0\) at fixed \(q_\rho\), the TM
amplitude becomes
\begin{equation}
 r_{\TM}(q_\rho,0)=
 \frac{(\varepsilon_1-\varepsilon_2)\varepsilon_0 q_\rho
       -\chi_{2\mathrm D}(q_\rho,0)}
      {(\varepsilon_1+\varepsilon_2)\varepsilon_0 q_\rho
       +\chi_{2\mathrm D}(q_\rho,0)}.
\label{eq:static_rTM}
\end{equation}
For a symmetric dielectric environment, \(\varepsilon_1=\varepsilon_2=\varepsilon_c\), this reduces to
\begin{equation}
 r_{\TM}(q_\rho,0)=
 -\frac{Q_{\mathrm{scr}}(q_\rho)}
 {q_\rho+Q_{\mathrm{scr}}(q_\rho)},
 \qquad
 Q_{\mathrm{scr}}(q_\rho)=
 \frac{\chi_{2\mathrm D}(q_\rho,0)}{\varepsilon_0\varepsilon_c}.
\label{eq:rTM_symmetric_static}
\end{equation}
Thus an ordinary charge-response sheet naturally approaches the
Dirichlet/PEC-like branch \(r_{\TM}\to -1\) when its static screening
response becomes large.  The opposite parity branch \(r_{\TM}\to +1\)
corresponds to a Neumann/PMC-like or high-impedance boundary for the
relevant scalar TM propagator.  This distinction between \(r_{\TM}\to -1\) and
\(r_{\TM}\to +1\) is central: the magnitude controls reflectivity, but the
phase/parity selects screening versus antiscreening.
A complete derivation of the material-agnostic sheet boundary conditions, together with their specialization to graphene, is given in Supplementary Information Sec.~S4; Supplementary Fig.~S3 summarizes the corresponding TM and TE scattering geometries and field discontinuities.

\section{Feynman, retarded, advanced, and vacuum propagators}
\label{sec:kernels}

The full QED-action derivation and the greater/lesser, Keldysh, retarded, advanced, Feynman, and symmetric-vacuum conventions are given in Supplementary Information Sec.~S1.  The associated retarded and Feynman pole placements are worked out in Supplementary Information Sec.~S2 and Appendix~\ref{app:contours}; Supplementary Fig.~S1 shows the positive-frequency \(q_z\)-plane contour used for the reduced retarded propagator.

\subsection{Photon propagators}

Let
\begin{equation}
\tau=t-t'.
\label{eq:relative_time}
\end{equation}
The standard greater and lesser bosonic propagators are
\begin{align}
D^{>\mu\nu}(x,x')
&\equiv
-\I\langle0|\hat A^\mu(x)\hat A^\nu(x')|0\rangle,
\label{eq:Dgreater_def}
\\
D^{<\mu\nu}(x,x')
&\equiv
-\I\langle0|\hat A^\nu(x')\hat A^\mu(x)|0\rangle.
\label{eq:Dless_def}
\end{align}
The same factor \(-\I\) appears in both definitions because
\(\hat A^\mu\) is a bosonic field.  The four-vector Feynman propagator is
then
\begin{align}
D_F^{\mu\nu}(x,x')
&\equiv
-\I\langle0|T\hat A^\mu(x)\hat A^\nu(x')|0\rangle
\nn\\
&=
\theta(\tau)D^{>\mu\nu}(x,x')
+
\theta(-\tau)D^{<\mu\nu}(x,x').
\label{eq:DF_A_def}
\end{align}
The retarded and advanced propagators are defined by
\begin{align}
D_R^{\mu\nu}(x,x')
&\equiv
-\I\theta(\tau)
\langle0|[\hat A^\mu(x),\hat A^\nu(x')]|0\rangle,
\label{eq:DR_def}
\\
D_A^{\mu\nu}(x,x')
&\equiv
+\I\theta(-\tau)
\langle0|[\hat A^\mu(x),\hat A^\nu(x')]|0\rangle.
\label{eq:DA_def}
\end{align}
These definitions imply
\begin{equation}
D_F^{\mu\nu}
=
D_R^{\mu\nu}+D^{<\mu\nu}
=
D_A^{\mu\nu}+D^{>\mu\nu}.
\label{eq:DF_greater_lesser_relations}
\end{equation}

The Keldysh propagator and the physical symmetric vacuum-correlation
propagator are
\begin{align}
D_K^{\mu\nu}(x,x')
&\equiv
D^{>\mu\nu}(x,x')+D^{<\mu\nu}(x,x')
\nn\\
&=
-\I\langle0|\{\hat A^\mu(x),\hat A^\nu(x')\}|0\rangle,
\label{eq:DK_def}
\\
D_{\rm vac}^{\mu\nu}(x,x')
&\equiv
\frac12
\langle0|\{\hat A^\mu(x),\hat A^\nu(x')\}|0\rangle
=
\frac{\I}{2}D_K^{\mu\nu}(x,x').
\label{eq:Dvac_def}
\end{align}
Using
\(\theta(\tau)=[1+\sgn(\tau)]/2\) and
\(\theta(-\tau)=[1-\sgn(\tau)]/2\), one obtains
\begin{align}
D_F^{\mu\nu}
&=
\frac12
\left[
D_R^{\mu\nu}
+
D_A^{\mu\nu}
+
D_K^{\mu\nu}
\right]
\nn\\
&=
\frac12
\left[
D_R^{\mu\nu}
+
D_A^{\mu\nu}
\right]
-
\I D_{\rm vac}^{\mu\nu}.
\label{eq:DF_split}
\end{align}
Thus the Feynman propagator contains the time-symmetric response propagator
\((D_R+D_A)/2\) and the symmetric vacuum-fluctuation propagator
\(D_{\rm vac}\).  In general, it is not equal to
\(D_R+\I D_{\rm vac}\).

For a homogeneous, isotropic, lossless, and nondispersive dielectric
medium \(j\), define
\begin{equation}
\Delta_j(q)
\equiv
\varepsilon_j\left(\frac{\omega}{c}\right)^2-q_\rho^2-q_z^2.
\label{eq:Delta_j_def}
\end{equation}
With the conventions above, the Lorenz-gauge momentum-space propagators are
\begin{align}
D_F^{\mu\nu}(q)
&=
-\frac{g^{\mu\nu}}{\Delta_j(q)+\I0^+},
\label{eq:DF_pole}
\\
D_R^{\mu\nu}(q)
&=
-\frac{g^{\mu\nu}}
{\varepsilon_j(\omega/c+\I0^+)^2-q_\rho^2-q_z^2}
\nn\\
&\equiv
-\frac{g^{\mu\nu}}
{\Delta_j(q)+\I0^+\sgn\omega},
\label{eq:DR_pole}
\\
D_A^{\mu\nu}(q)
&=
-\frac{g^{\mu\nu}}
{\varepsilon_j(\omega/c-\I0^+)^2-q_\rho^2-q_z^2}
\nn\\
&\equiv
-\frac{g^{\mu\nu}}
{\Delta_j(q)-\I0^+\sgn\omega}.
\label{eq:DA_pole}
\end{align}
The overall minus sign changes the residues relative to the previous
normalization but does not change any pole location.  At zero temperature,
\begin{equation}
D_{\rm vac}^{\mu\nu}(q)
=
-\pi g^{\mu\nu}\delta[\Delta_j(q)].
\label{eq:Dvac_mom}
\end{equation}
The symmetric vacuum propagator is even in frequency; the factor
\(\sgn\omega\) belongs instead to the commutator or spectral propagator.  In
particular,
\begin{align}
C^{\mu\nu}(q)
&\equiv
\langle0|[\hat A^\mu,\hat A^\nu]|0\rangle_q
\nn\\
&=
\I\left[
D_R^{\mu\nu}(q)-D_A^{\mu\nu}(q)
\right]
=
-2\Im D_R^{\mu\nu}(q),
\label{eq:commutator_spectral_identity}
\\
D_{\rm vac}^{\mu\nu}(q)
&=
-\sgn\omega\,\Im D_R^{\mu\nu}(q).
\label{eq:Dvac_ImDR}
\end{align}
The detailed greater/lesser and finite-temperature derivations are given
in the Supplementary Information.

\subsection{Electric-field propagators and Maxwell Green tensor}

For optical emission and FRET, the relevant operators are the electric
fields.  We define
\begin{align}
\calD^{EE}_{F,ij}(x,x')
&\equiv
-\I\langle0|T\hat E_i(x)\hat E_j(x')|0\rangle,
\label{eq:DEE_F_def}
\\
\calD^{EE}_{R,ij}(x,x')
&\equiv
-\I\theta(t-t')
\langle0|[\hat E_i(x),\hat E_j(x')]|0\rangle,
\label{eq:DEE_R_def}
\\
\calD^{EE}_{A,ij}(x,x')
&\equiv
+\I\theta(t'-t)
\langle0|[\hat E_i(x),\hat E_j(x')]|0\rangle.
\label{eq:DEE_A_def}
\end{align}
The response-normalized retarded electric susceptibility is
\begin{align}
\chi^{EE}_{R,ij}(x,x')
&\equiv
-\frac{1}{\hbar}\calD^{EE}_{R,ij}(x,x')
\nn\\
&=
\frac{\I}{\hbar}\theta(t-t')
\langle0|[\hat E_i(x),\hat E_j(x')]|0\rangle.
\label{eq:chi_EE_R_def}
\end{align}
In frequency space it is related to the retarded Maxwell dyadic by
\begin{align}
\chi^{EE}_{R,ij}(\bx,\bx';\omega)
&=
\mu_0\omega^2G_{R,ij}(\bx,\bx';\omega),
\nn\\
\calD^{EE}_{R,ij}
&=
-\hbar\mu_0\omega^2G_{R,ij}.
\label{eq:D_EE_GR}
\end{align}
The retarded Maxwell dyadic satisfies
\begin{equation}
\left[
\nabla\times\nabla\times
-
\frac{(\omega+\I0^+)^2}{c^2}\varepsilon(\bx,\omega)
\right]
\obG_R(\bx,\bx';\omega)
=
\bI\,\delta(\bx-\bx').
\label{eq:Maxwell_GR}
\end{equation}
For reciprocal media,
\begin{equation}
\obG_A(\bx,\bx';\omega)
=
\left[
\obG_R(\bx',\bx;\omega)
\right]^\dagger.
\label{eq:Maxwell_GA_reciprocity}
\end{equation}
Consequently,
\begin{equation}
\calD^{EE}_{R}-\calD^{EE}_{A}
=
-2\I\hbar\mu_0\omega^2\Im\obG_R.
\label{eq:DEE_spectral_difference}
\end{equation}

The physical positive-frequency vacuum correlation does not inherit the
convention-dependent overall sign of the operator propagator.  At zero
temperature, macroscopic QED gives
\begin{align}
&\left\langle0\left|
\hat E_i^{(+)}(\bx,\omega)
\hat E_j^{(-)}(\bx',\omega')
\right|0\right\rangle
\nn\\
&\qquad
=
\frac{\hbar\mu_0}{\pi}\omega^2
\Im G_{R,ij}(\bx,\bx';\omega)
\delta(\omega-\omega')\theta(\omega).
\label{eq:mqed_correlator}
\end{align}
Equation~\eqref{eq:mqed_correlator} is the vacuum
fluctuation-dissipation relation and provides the operational connection
between vacuum fluctuations and the retarded Maxwell Green tensor.

\section{Reduced retarded Green function in the two-sheet cavity}
\label{sec:cavitygreen}

\subsection{Weyl representation}

The planar geometry allows a Weyl decomposition.  For any scalar or
polarization-resolved component,
\begin{equation}
 \calD_{\text R}(x_\rho,x_z,x_z';\omega)
 =
 \frac{1}{2\pi}
 \int_0^{\infty}
 q_\rho\,\dd q_\rho\,
 J_0(q_\rho x_\rho)\,
 \tilde\calD_{\text R}(q_\rho;x_z,x_z';\omega).
\label{eq:Weyl_scalar}
\end{equation}
The electromagnetic dyadic is obtained by multiplying the reduced scalar
propagator by the TE/TM polarization projectors before the angular
integration.  In the near-field TM limit, the scalar representation captures
the universal distance dependence; tensor factors determine
orientation-dependent prefactors.

For a homogeneous spacer, the reduced free retarded scalar propagator is
\begin{equation}
 \tilde D_{\text R}^{(0)}(q_\rho;x_z,x_z';\omega)
 =
 \frac{\I}{2Q_c}
 e^{\I Q_c|x_z-x_z'|},
\label{eq:free_reduced_retarded}
\end{equation}
with \(Q_c\) defined in Eq.~\eqref{eq:Qc_real_def}.  For evanescent
components, \(Q_c=\I\bar Q_c\), one obtains
\begin{equation}
 \tilde D_{\text R}^{(0)}(q_\rho;x_z,x_z';\omega)
 =
 \frac{1}{2\bar Q_c}
 e^{-\bar Q_c|x_z-x_z'|}.
\label{eq:free_reduced_retarded_evanescent}
\end{equation}

\subsection{Multiple-reflection resummation}

Let the lower and upper reflection amplitudes for polarization \(\lambda\)
be \(r^-_{\lambda}\) and \(r^+_{\lambda}\).  Summing all paths that
alternate between the two interfaces gives
\begin{align}
\tilde\calD_{\lambda,\text R}
(q_\rho;x_z,x_z';\omega)
&=
\frac{\I}{2Q_c}
\Bigg[
e^{\I Q_c|x_z-x_z'|}
\nn\\
&+
\frac{
r^-_{\lambda}e^{\I Q_c(x_z+x_z'+W)}
+
r^+_{\lambda}e^{\I Q_c(W-x_z-x_z')}
}
{1-r^+_{\lambda}r^-_{\lambda}e^{2\I Q_c W}}
\nn\\
&+
\frac{
2r^+_{\lambda}r^-_{\lambda}e^{2\I Q_cW}
\cos[Q_c(x_z-x_z')]
}
{1-r^+_{\lambda}r^-_{\lambda}e^{2\I Q_cW}}
\Bigg].
\label{eq:general_cavity_reduced}
\end{align}
This is the one-dimensional version of the full multiple-scattering series:
the denominator is the round-trip factor and the numerator accounts for
direct, one-reflection, and completed-round-trip paths.
The complete path-by-path derivation is given in Supplementary Information Sec.~S2, and Supplementary Fig.~S2 illustrates the direct, single-reflection, and repeated round-trip contributions that generate the Fabry--P'erot denominator.

For identical interfaces,
\(r^+_{\lambda}=r^-_{\lambda}=r_{\lambda}\), and midplane sources
\(x_z=x_z'=0\), Eq.~\eqref{eq:general_cavity_reduced} collapses to
\begin{equation}
 \tilde\calD_{\lambda,\text R}(q_\rho;0,0;\omega)
 =
 \frac{\I}{2Q_c}
 \frac{1+r_{\lambda}(q_\rho,\omega)e^{\I Q_c W}}
 {1-r_{\lambda}(q_\rho,\omega)e^{\I Q_c W}}.
\label{eq:midplane_reduced_retarded}
\end{equation}
For evanescent components this becomes
\begin{equation}
 \tilde\calD_{\lambda,\text R}(q_\rho;0,0;\omega)
 =
 \frac{1}{2\bar Q_c}
 \frac{1+r_{\lambda}(q_\rho,\omega)e^{-\bar Q_c W}}
 {1-r_{\lambda}(q_\rho,\omega)e^{-\bar Q_c W}}.
\label{eq:midplane_evanescent}
\end{equation}
At imaginary frequency,
\begin{equation}
 \tilde\calD_{\lambda}(q_\rho;0,0;\I\xi)
 =
 \frac{1}{2\bar Q_c(q_\rho,\xi)}
 \frac{1+r_{\lambda}(q_\rho,\I\xi)e^{-\bar Q_c W}}
 {1-r_{\lambda}(q_\rho,\I\xi)e^{-\bar Q_c W}},
\label{eq:midplane_imag}
\end{equation}
up to the overall Euclidean sign convention.  These equations are the
central reduction: all geometry is contained in a universal denominator,
and all microscopic material physics enters through \(r_{\lambda}\).

\subsection{Static TM image lattice}

In the near-field electrostatic limit, \(\omega\to0\) and
\(\bar Q_c\to q_\rho\).  The TM channel dominates, and
Eq.~\eqref{eq:midplane_evanescent} gives
\begin{equation}
 D_{\mathrm{stat}}(x_\rho)
 =
 \frac{1}{4\pi\varepsilon_0\varepsilon_c}
 \int_0^{\infty}\dd q_\rho\,
 J_0(q_\rho x_\rho)
 \frac{1+r_{\TM}e^{-q_\rho W}}
 {1-r_{\TM}e^{-q_\rho W}}.
\label{eq:static_integral}
\end{equation}
Expanding the geometric series yields
\begin{equation}
 \frac{1+r e^{-q_\rho W}}{1-r e^{-q_\rho W}}
 =
 1+2\sum_{m=1}^{\infty}r^m e^{-m q_\rho W},
\end{equation}
so that
\begin{equation}
 D_{\mathrm{stat}}(x_\rho)
 =
 \frac{1}{4\pi\varepsilon_0\varepsilon_c}
 \sum_{m=-\infty}^{\infty}
 \frac{r_{\TM}^{|m|}}{\sqrt{x_\rho^2+m^2W^2}}.
\label{eq:static_image_lattice}
\end{equation}
The sign of \(r_{\TM}\) is now the image parity.  Negative \(r_{\TM}\) gives
alternating-sign images and positive \(r_{\TM}\) gives same-sign images.
The signed image construction, its dyadic derivatives, and its Poisson/Bessel resummation are derived in Supplementary Information Sec.~S7 and Appendix~\ref{app:poisson}; Supplementary Fig.~S5 visualizes the alternating-sign and same-sign image lattices associated with the two parity branches.

It is useful to introduce
\begin{equation}
 \eta=\frac{x_\rho}{W},
 \qquad
 S(\eta;r)=
 \sum_{m=-\infty}^{\infty}
 \frac{r^{|m|}}{\sqrt{\eta^2+m^2}},
\end{equation}
so that
\begin{equation}
 D_{\mathrm{stat}}(x_\rho)
 =
 \frac{1}{4\pi\varepsilon_0\varepsilon_c W}
 S(\eta;r_{\TM}).
\end{equation}
The two perfect-reflector parity branches are
\begin{align}
S(\eta;-1)
&=
2\sum_{\ell=0}^{\infty}K_0[(2\ell+1)\pi\eta],
\label{eq:S_minus_Bessel}\\
S(\eta;+1)
&=
\frac{1}{\eta}
+
2\sum_{\ell=1}^{\infty}K_0(2\pi\ell\eta).
\label{eq:S_plus_Bessel}
\end{align}
Thus
\begin{equation}
D_{\mathrm{stat}}^{(-)}(x_\rho)
\simeq
\frac{1}{2\pi\varepsilon_0\varepsilon_c W}
K_0(\pi x_\rho/W)
\propto
\frac{e^{-\pi x_\rho/W}}{\sqrt{x_\rho/W}}
\label{eq:static_screening_asymptotic}
\end{equation}
for \(r_{\TM}\to-1\) and \(x_\rho\gg W\).  Conversely, for
\(r_{\TM}=1-\delta\) with \(0<\delta\ll1\), let
\begin{equation}
 a\equiv -\ln r_{\TM}\simeq \delta,
 \qquad
 x_{\rho,*}\sim \frac{W}{a}\simeq \frac{W}{1-r_{\TM}}.
\end{equation}
The nearly gapless \(\ell=0\) mode gives
\begin{equation}
D_{\mathrm{stat}}^{(+)}(x_\rho)
\simeq
\frac{1}{2\pi\varepsilon_0\varepsilon_c W}
\left[
\ln\left(\frac{2W}{a x_\rho}\right)-\gamma_E
\right],
\label{eq:log_antiscreening}
\end{equation}
with $W\ll x_\rho\ll x_{\rho,*}$.
This is the quasi-two-dimensional logarithmic antiscreening regime.

\section{Hybridization and self-energy interpretation of the retarded Green function}
\label{sec:hybridization_self_energy_GR}

A more detailed derivation of the same construction, including the quadratic hybridization action, integration over the sheet degrees of freedom, Dyson equation, single-interface \(T\)-matrix, reflection-amplitude mapping, hybridized cavity poles, and spectral representation, is given in Supplementary Information Sec.~S3.

The reflection-amplitude formulation used throughout this work can also be
viewed as a self-energy description.  In this complementary picture, the
photon modes of the center dielectric spacer hybridize with electromagnetic
and charge-current modes supported by the top and bottom conducting sheets.
Integrating out these sheet degrees of freedom produces a retarded
self-energy localized at the interfaces.  The full retarded Green function
in the spacer is then obtained from a Dyson equation.  The Fresnel reflection
amplitudes \(r_{\rm TM/TE}(q_\rho,\omega)\) are the corresponding on-shell
surface scattering amplitudes.

The electrically controlled Er-graphene decay observed by Cano \textit{et al.} provides a direct experimental realization of this self-energy viewpoint \cite{Cano2020ErGraphene}.  In the TM channel, interband electron-hole excitations and intraband/plasmon emission contribute to the absorptive part of the graphene-induced retarded self-energy and hence to irreversible Er decay.  The same graphene response also produces a principal-value, reactive contribution to the self-energy, which becomes the dominant effect for sufficiently off-resonant or in-gap virtual-plasmon coupling.

We first write the reduced retarded Green function in the center dielectric
medium, before coupling to the conducting sheets, as
\begin{equation}
\tilde D_{\rm R}^{(c)}(q_\rho;x_z,x_z';\omega)
=
\frac{\I}{2Q_c}
e^{\I Q_c|x_z-x_z'|},
\label{eq:bare_reduced_GR_center}
\end{equation}
where
\begin{equation}
Q_c(q_\rho,\omega)
=
\left[
\varepsilon_c\left(\frac{\omega}{c}+\I0^+\right)^2-q_\rho^2
\right]^{1/2},
\qquad
\Im Q_c\ge0 .
\label{eq:Qc_retarded_self_energy}
\end{equation}
For evanescent components \(Q_c=\I\bar Q_c\), with
\(\bar Q_c>0\), Eq.~\eqref{eq:bare_reduced_GR_center} becomes the usual
exponentially decaying propagator
\begin{equation}
\tilde D_{\rm R}^{(c)}(q_\rho;x_z,x_z';\omega)
=
\frac{1}{2\bar Q_c}
e^{-\bar Q_c|x_z-x_z'|}.
\label{eq:bare_reduced_GR_center_evanescent}
\end{equation}

The two conducting sheets are located at
\[
x_z^\pm=\pm W/2.
\]
Their internal electromagnetic, plasmonic, or particle-hole degrees of
freedom can be integrated out.  The result is a retarded surface self-energy
of the form
\begin{align}
\Sigma_{\rm R}^{(\sigma)}(q_\rho,\omega;x_z,x_z')
&=
\sum_{\zeta=\pm}
\Sigma_{\zeta}^{(\sigma)}(q_\rho,\omega)\,
\delta(x_z-x_z^\zeta)\,
\delta(x_z'-x_z^\zeta),
\label{eq:surface_self_energy_local}
\end{align}
where \(\sigma=\mathrm{TM},\mathrm{TE}\).  Microscopically,
\(\Sigma_{\zeta}^{(\sigma)}\) is determined by the appropriate retarded
surface response function of sheet \(\zeta\).  In the TM sector it is tied
to the longitudinal charge-density response, or equivalently to the
longitudinal sheet conductivity.  In the TE sector it is tied to the
transverse current response.

The full reduced retarded Green function obeys the Dyson equation
\begin{align}
\tilde{\calD}_{\rm R}^{(\sigma)}
(q_\rho;x_z,x_z';\omega)
&=
\tilde D_{\rm R}^{(c)}
(q_\rho;x_z,x_z';\omega)
\nn\\
&+
\int dx_1 dx_2\,
\tilde D_{\rm R}^{(c)}
(q_\rho;x_z,x_1;\omega)
\nn\\
&\times
\Sigma_{\rm R}^{(\sigma)}
(q_\rho,\omega;x_1,x_2)
\tilde{\calD}_{\rm R}^{(\sigma)}
(q_\rho;x_2,x_z';\omega).
\label{eq:Dyson_reduced_GR}
\end{align}
Because the self-energy is localized at the two interfaces, the Dyson
equation reduces to a finite-dimensional matrix problem in the surface
subspace \(x_z=x_z^\pm\).

It is useful to introduce the single-interface retarded \(T\)-matrix,
\begin{equation}
T_{\zeta}^{(\sigma)}(q_\rho,\omega)
=
\frac{
\Sigma_{\zeta}^{(\sigma)}(q_\rho,\omega)
}{
1-\Sigma_{\zeta}^{(\sigma)}(q_\rho,\omega)\,
\tilde D_{\rm R}^{(c)}(q_\rho;x_z^\zeta,x_z^\zeta;\omega)
}.
\label{eq:Tmatrix_surface_def}
\end{equation}
Using
\begin{equation}
\tilde D_{\rm R}^{(c)}(q_\rho;x_z^\zeta,x_z^\zeta;\omega)
=
\frac{\I}{2Q_c},
\label{eq:onsite_D0_surface}
\end{equation}
we obtain
\begin{equation}
T_{\zeta}^{(\sigma)}
=
\frac{
\Sigma_{\zeta}^{(\sigma)}
}{
1-\dfrac{\I\Sigma_{\zeta}^{(\sigma)}}{2Q_c}
}.
\label{eq:Tmatrix_surface_explicit}
\end{equation}
The reflection amplitude is the on-shell surface scattering amplitude
normalized by the free outgoing propagation factor.  With the convention of
Eq.~\eqref{eq:bare_reduced_GR_center}, this gives
\begin{equation}
T_{\zeta}^{(\sigma)}
=
-2\I Q_c\,r_{\zeta}^{(\sigma)}.
\label{eq:Tmatrix_reflection_relation}
\end{equation}
Equivalently,
\begin{equation}
r_{\zeta}^{(\sigma)}
=
-\frac{
\Sigma_{\zeta}^{(\sigma)}
}{
2\I Q_c+\Sigma_{\zeta}^{(\sigma)}
},
\qquad
\Sigma_{\zeta}^{(\sigma)}
=
-\frac{
2\I Q_c\,r_{\zeta}^{(\sigma)}
}{
1+r_{\zeta}^{(\sigma)}
}.
\label{eq:self_energy_reflection_relation}
\end{equation}
Equation~\eqref{eq:self_energy_reflection_relation} is the desired connection
between the self-energy and reflection-amplitude descriptions.  A transparent
sheet has \(\Sigma_{\zeta}^{(\sigma)}=0\) and hence
\(r_{\zeta}^{(\sigma)}=0\).  A strongly reflecting Dirichlet/PEC-like
boundary corresponds to \(r_{\zeta}^{(\sigma)}\to -1\), which is obtained
from a divergent surface self-energy.  The opposite Neumann/PMC-like parity
branch corresponds to
\(\Sigma_{\zeta}^{(\sigma)}\to-\I Q_c\), which produces
\(r_{\zeta}^{(\sigma)}\to +1\) in the relevant polarization channel.

This relation can also be derived from a microscopic hybridization model.
Let \(a_{q_\rho}\) denote a photon mode in the center dielectric spacer and
let \(b_{\zeta,\lambda,q_\rho}\) denote a surface electromagnetic or plasmonic
mode of sheet \(\zeta\), with branch index \(\lambda\).  A minimal quadratic
Hamiltonian has the form
\begin{align}
H_{\rm hyb}
&=
\sum_{q_\rho}
\hbar\omega_c(q_\rho)
a_{q_\rho}^{\dagger}a_{q_\rho}
+
\sum_{\zeta,\lambda,q_\rho}
\hbar\Omega_{\zeta\lambda}(q_\rho)
b_{\zeta,\lambda,q_\rho}^{\dagger}
b_{\zeta,\lambda,q_\rho}
\nn\\
&+
\sum_{\zeta,\lambda,q_\rho}
\left[
g_{\zeta\lambda}(q_\rho)
a_{q_\rho}^{\dagger}
b_{\zeta,\lambda,q_\rho}
+
g_{\zeta\lambda}^{*}(q_\rho)
b_{\zeta,\lambda,q_\rho}^{\dagger}
a_{q_\rho}
\right].
\label{eq:hybridization_model}
\end{align}
Integrating out the sheet modes gives the retarded self-energy
\begin{equation}
\Sigma_{\zeta}^{(\sigma)}(q_\rho,\omega)
=
\sum_{\lambda}
\frac{
|g_{\zeta\lambda}^{(\sigma)}(q_\rho)|^2
}{
\omega-\Omega_{\zeta\lambda}(q_\rho)+\I0^+
}
+
\Sigma_{\zeta,\rm cont}^{(\sigma)}(q_\rho,\omega),
\label{eq:microscopic_self_energy_modes}
\end{equation}
where \(\Sigma_{\zeta,\rm cont}^{(\sigma)}\) denotes the corresponding
continuum contribution from particle-hole excitations, dissipative sheet
currents, and radiative channels.  In a macroscopic description the detailed
mode sum in Eq.~\eqref{eq:microscopic_self_energy_modes} is replaced by the
measured or calculated sheet response function, and hence by the Fresnel
coefficient \(r_{\zeta}^{(\sigma)}(q_\rho,\omega)\).

The equivalence between the hybridization picture and the reflection picture
becomes especially transparent for two interfaces.  A wave that leaves a
point \(x_z'\), scatters from interface \(\zeta\), and returns to a point
\(x_z\) contributes
\begin{align}
\tilde D_{\rm R}^{(c)}(x_z,x_z^\zeta)\,
T_{\zeta}^{(\sigma)}\,
\tilde D_{\rm R}^{(c)}(x_z^\zeta,x_z')
=
\left(
\frac{\I}{2Q_c}
\right)
e^{\I Q_c|x_z-x_z^\zeta|}
\nn\\
\times
\left(-2\I Q_c r_{\zeta}^{(\sigma)}\right)
\left(
\frac{\I}{2Q_c}
\right)
e^{\I Q_c|x_z^\zeta-x_z'|}
\nn\\
=
\frac{\I}{2Q_c}\,
r_{\zeta}^{(\sigma)}
e^{\I Q_c\left(|x_z-x_z^\zeta|+|x_z^\zeta-x_z'|\right)}.
\label{eq:single_reflection_from_T}
\end{align}
Thus each surface \(T\)-matrix insertion is exactly equivalent to multiplying
by the reflection amplitude \(r_{\zeta}^{(\sigma)}\), together with the
appropriate free propagation factors.

Likewise, a complete round trip between the two interfaces gives
\begin{align}
&
\tilde D_{\rm R}^{(c)}(x_z^+,x_z^-)\,
T_{-}^{(\sigma)}\,
\tilde D_{\rm R}^{(c)}(x_z^-,x_z^+)\,
T_{+}^{(\sigma)}
\nn\\
&\qquad
=
r_{+}^{(\sigma)}r_{-}^{(\sigma)}
e^{2\I Q_cW}.
\label{eq:round_trip_from_T}
\end{align}
Therefore summing all repeated hybridization events with the top and bottom
sheets produces the same Fabry-Pérot denominator as the multiple-reflection
approach,
\begin{equation}
\tilde{\calR}_{\rm R}^{(\sigma)}
=
\left[
1-r_{+}^{(\sigma)}r_{-}^{(\sigma)}
e^{2\I Q_cW}
\right]^{-1}.
\label{eq:self_energy_FP_denominator}
\end{equation}
The poles of the dressed retarded Green function are consequently determined
by
\begin{equation}
1-r_{+}^{(\sigma)}(q_\rho,\omega)
r_{-}^{(\sigma)}(q_\rho,\omega)
e^{2\I Q_c(q_\rho,\omega)W}
=0.
\label{eq:hybridized_mode_condition}
\end{equation}
These poles are the hybrid photon-surface-mode resonances of the
conductor-dielectric-conductor structure.  In the weak-coupling limit,
they are only small shifts and broadenings of the photon modes of the
central spacer.  In the strong-coupling limit, they become genuine
hybrid light-matter modes with appreciable weight in both the dielectric
spacer and the conducting sheets.

This self-energy viewpoint clarifies the role of the reflection coefficient:
\(r_{\rm TM/TE}\) is not an additional phenomenological assumption, but the
on-shell representation of the sheet-induced photon self-energy.  The
material-agnostic nature of our theory follows from the fact that all
microscopic information about the conducting sheets enters the spacer Green
function through the retarded surface self-energies, or equivalently through
the measurable reflection amplitudes \(r_{\rm TM/TE}(q_\rho,\omega)\).
Consequently, the same hybridization mechanism controls spontaneous emission
and FRET.  The donor excites a retarded photon field in the center dielectric
medium; this field hybridizes with the top and bottom conducting layers
through \(\Sigma_{\zeta}^{(\sigma)}\); the resulting dressed Green tensor
\(\obG_{\rm R}\) then determines both the local density of states for
spontaneous emission and the donor-acceptor propagation factor entering
FRET.

\section{Spontaneous emission from the vacuum propagator}\label{sec:spontaneous}

The corresponding vacuum-correlation derivation is given in Supplementary Information Sec.~S5, while finite-temperature and absorbing-medium generalizations are summarized in Appendix~\ref{app:finiteT}.

Consider a two-level emitter at position $\bx_0$ with transition frequency $\omega_0$ and transition dipole $\bd$.  In the dipole gauge,
\begin{equation}
 \hat H_I(t)=-\hat{\bd}(t)\cdot\hat{\bE}(\bx_0,t).
\end{equation}
For initial state $|i\rangle=|e,0_\gamma\rangle$ and final states $|f\rangle=|g,1_{\lambda}\rangle$, Fermi's golden rule gives
\begin{equation}
 \Gamma=\frac{2\pi}{\hbar}\sum_{\lambda}
 \left|\langle g,1_{\lambda}|\hat H_I(0)|e,0_\gamma\rangle\right|^2
 \delta(\omega_{\lambda}-\omega_0).
\end{equation}
The mode sum can be written as a vacuum field correlator,
\begin{equation}
 \Gamma=\frac{2}{\hbar^2}\int_{-\infty}^{\infty}\dd t\,
 e^{\I\omega_0t}d_i d_j^*
 \left\langle0\left|\hat E_i^{(+)}(\bx_0,t)\hat E_j^{(-)}(\bx_0,0)\right|0\right\rangle.
\label{eq:Gamma_corr}
\end{equation}
Using Eq.~\eqref{eq:mqed_correlator} gives the standard LDOS formula
\begin{equation}
 \Gamma(\bx_0,\omega_0)
 =\frac{2\mu_0\omega_0^2}{\hbar}
 d_i d_j^*\Im G_{\text R,ij}(\bx_0,\bx_0;\omega_0).
\label{eq:sp_emission_rate}
\end{equation}
Equivalently, since the vacuum propagator is the absorptive part of the retarded propagator,
\begin{equation}
 \Gamma=\frac{2\omega_0^2}{\hbar^2}d_i d_j^*D^{(+)}_{\mathrm{vac},ij}(\bx_0,\bx_0;\omega_0).
\end{equation}
This makes the division of roles transparent: the retarded Green tensor describes the density of outgoing modes, while the vacuum propagator provides the nonzero zero-point correlation that allows decay from the photon vacuum.

In homogeneous free space,
\begin{equation}
 \Im G_{\text R,ij,0}(\bx_0,\bx_0;\omega_0)=\frac{\omega_0}{6\pi c}\delta_{ij},
\end{equation}
so Eq.~\eqref{eq:sp_emission_rate} reduces to
\begin{equation}
 \Gamma_0=\frac{\omega_0^3|\bd|^2}{3\pi\varepsilon_0\hbar c^3}.
\end{equation}
In the two-sheet geometry the Purcell factor is
\begin{equation}
 F_P(\bx_0,\omega_0)=\frac{\Gamma}{\Gamma_0}
 =\frac{d_i d_j^*\Im G_{\text R,ij}(\bx_0,\bx_0;\omega_0)}{|\bd|^2\omega_0/(6\pi c)}.
\label{eq:Purcell_factor}
\end{equation}
Because $\obG_{\text R}$ is built from the same reduced denominator in Eq.~\eqref{eq:midplane_reduced_retarded}, spontaneous emission is tunable by the same reflection amplitudes $r_{\TM/\TE}(q,\omega_0)$.  The important distinction from static Coulomb tuning is that spontaneous emission samples the \emph{on-shell spectral density}: propagating modes, lossy guided modes, plasmonic modes, and radiative continuum states all contribute through $\Im\,\obG_{\text R}(\bx_0,\bx_0;\omega_0)$.

\section{FRET from the retarded Green tensor}\label{sec:fret}

The full second-order transition-operator derivation is given in Supplementary Information Sec.~S6 and Appendix~\ref{app:Tmatrix}.  Supplementary Fig.~S4 displays the two vacuum time orderings whose mode sums combine into the full retarded donor-to-acceptor Green tensor.

\subsection{Second-order transition amplitude}

The donor-acceptor system has initial and final states
\begin{align}
 |i\rangle&=|e_D,g_A,0_\gamma\rangle,\\
 |f\rangle&=|g_D,f_A,0_\gamma\rangle,
\end{align}
where $|f_A\rangle$ denotes an acceptor excited state or vibronic manifold.  The interaction Hamiltonian is
\begin{equation}
 \hat H_I=-\hat{\bd}_D\cdot\hat{\bE}(\bx_D)-\hat{\bd}_A\cdot\hat{\bE}(\bx_A).
\end{equation}
The second-order $T$-matrix is
\begin{equation}
 M^{(2)}_{fi}=\left\langle f\left|\hat H_I\frac{1}{E_i-\hat H_0+\I0^+}\hat H_I\right|i\right\rangle.
\label{eq:Tmatrix_second_order}
\end{equation}
Inserting intermediate one-photon states and keeping both time orderings gives
\begin{align}
M^{(2)}_{fi}
&=\frac{1}{2\varepsilon_0}\sum_{\lambda}\omega_{\lambda}
 d_{A,i}^{(f)}d_{D,j}
 \bigg[
 \frac{f_{\lambda,i}(\bx_A)f_{\lambda,j}^{*}(\bx_D)}{\omega_D-\omega_{\lambda}+\I0^+}
\nn\\ 
 &-\frac{f^{*}_{\lambda,i}(\bx_A)f_{\lambda,j}(\bx_D)}{\omega_D+\omega_{\lambda}-\I0^+}
 \bigg],
\label{eq:mode_sum_FRET}
\end{align}
where $d_{D,j}=\langle g_D|\hat d_{D,j}|e_D\rangle$ and $d^{(f)}_{A,i}=\langle f_A|\hat d_{A,i}|g_A\rangle$.  The two denominators are the two time orderings of virtual-photon exchange.

Using the spectral representation of the retarded Maxwell dyadic,
\begin{equation}
 G_{\text R,ij}(\bx,\bx';\omega)=\sum_{\lambda}\frac{c^2 f_{\lambda,i}(\bx)f_{\lambda,j}^*(\bx')}{\omega_{\lambda}^2-(\omega+\I0^+)^2},
\label{eq:spectral_GR}
\end{equation}
Eq.~\eqref{eq:mode_sum_FRET} reorganizes into
\begin{equation}
 M^{(2)}_{fi}=-\mu_0\omega_D^2\,
 d_{A,i}^{(f)}G_{\text R,ij}(\bx_A,\bx_D;\omega_D)d_{D,j},
\label{eq:FRET_amplitude_GR}
\end{equation}
for $\bx_A\neq\bx_D$; the contact term from mode completeness is absent for spatially separated donor and acceptor.  This equation is the central result of the propagator derivation: apart from the convention-dependent overall minus sign, the time-ordered second-order QED amplitude is the retarded electromagnetic response from donor to acceptor.  The minus sign cancels from the FRET rate.

\subsection{Golden-rule rate and acceptor absorption}

The FRET rate is
\begin{equation}
 \Gamma_{D\to A}=\frac{2\pi}{\hbar}\sum_{f_A}|M_{fi}^{(2)}|^2
 \delta(\omega_D-\omega_{f_Ag_A}).
\label{eq:FRET_GR_start}
\end{equation}
The acceptor final-state sum gives the absorptive part of the acceptor polarizability,
\begin{equation}
 \Im\,\alpha^A_{\ell m}(\omega)
 =\frac{\pi}{\hbar}\sum_{f_A}
 d^{(f)}_{A,\ell}d^{(f)*}_{A,m}
 \delta(\omega-\omega_{f_Ag_A}).
\label{eq:Im_alpha_acceptor}
\end{equation}
Substituting Eqs.~\eqref{eq:FRET_amplitude_GR} and \eqref{eq:Im_alpha_acceptor} into Eq.~\eqref{eq:FRET_GR_start} gives
\begin{align}
\Gamma_{D\to A}
=&\frac{2}{\hbar}(\mu_0\omega_D^2)^2
 d_{D,i}d_{D,j}^{*}
 G_{\text R,i\ell}(\bx_D,\bx_A;\omega_D)
\nn\\
\times &
\Im\,\alpha^A_{\ell m}(\omega_D)
 G_{\text A,mj}(\bx_A,\bx_D;\omega_D).
\label{eq:FRET_rate_general}
\end{align}
For an isotropic acceptor, $\Im\,\alpha^A_{\ell m}=\Im\,\alpha_A\delta_{\ell m}$, this reduces to
\begin{equation}
 \Gamma_{D\to A}=\frac{2}{\hbar}(\mu_0\omega_D^2)^2
 \Im\,\alpha_A(\omega_D)
 \left|\obG_{\text R}(\bx_A,\bx_D;\omega_D)\cdot\bd_D\right|^2.
\label{eq:FRET_rate_iso}
\end{equation}
This form is manifestly positive and explicitly separates three ingredients: donor oscillator strength, acceptor absorption, and nonlocal propagation through the structured environment.
For broadened donor and acceptor transitions, Supplementary Information Sec.~S6 gives the spectral-overlap form of the rate, while Sec.~S11 develops a semi-empirical multi-resonance donor/acceptor model; Supplementary Fig.~S6 illustrates the resulting spectral-overlap factor.

\subsection{Near-field bulk limit}

For $k_cR\ll1$ in a homogeneous dielectric, the retarded Green tensor reduces to the quasistatic dipole tensor,
\begin{equation}
 \mu_0\omega_D^2\obG_{\text R}(\bx_A,\bx_D;\omega_D)
 \to \frac{1}{4\pi\varepsilon_0\varepsilon_c}
 \frac{3\hat{\mathbf R}\hat{\mathbf R}-\mathbf{I}}{R^3},
\end{equation}
where $\mathbf R=\bx_A-\bx_D$.  Equation~\eqref{eq:FRET_rate_general} therefore gives
\begin{equation}
 \Gamma_{D\to A}^{(0)}\propto \frac{\kappa_{DA}^2}{R^6},
\end{equation}
where $\kappa_{DA}$ is the usual orientation factor.  In the two-sheet geometry, this bulk law is recovered when $r_{\TM}\to0$.

\section{Universal FRET tuning laws}\label{sec:tuning}

Supplementary Information Sec.~S8 gives the detailed transparent, Dirichlet/PEC, and Neumann/PMC-like asymptotics, building on the near-field dyadic and image-lattice derivations of Sec.~S7.  Orientation-dependent dyadic derivatives are collected independently in Appendix~\ref{app:dyadic}.

The FRET rate in Eq.~\eqref{eq:FRET_rate_general} is quadratic in the nonlocal Green tensor.  Therefore any cavity-induced change in the donor-acceptor coupling is squared in the transfer rate.  In the near-field TM-dominated regime, the relevant tensor is obtained by differentiating the scalar propagator,
\begin{align}
 \mathcal V_{DA}(x_\rho,\omega_D)
 &\equiv -M_{fi}^{(2)}
 =\mu_0\omega_D^2 d_{A,i}G_{\text R,ij}(\bx_A,\bx_D;\omega_D)d_{D,j}
 \nn\\
& \simeq d_{A,i}\partial_i\partial_jD_{\mathrm{stat}}(x_\rho)d_{D,j},
\label{eq:near_field_VDA}
\end{align}
with the understanding that for finite frequency $D_{\mathrm{stat}}$ is replaced by its retarded, frequency-dependent version.  The rate scales as
\begin{equation}
 \Gamma_{D\to A}\propto |\mathcal V_{DA}|^2\Im\,\alpha_A(\omega_D).
\end{equation}

\subsection{Transparent ON state: bulk power law}

For $r_{\TM}\to0$, Eq.~\eqref{eq:static_image_lattice} keeps only the $m=0$ image:
\begin{equation}
 D_{\mathrm{ON}}(x_\rho)=\frac{1}{4\pi\varepsilon_0\varepsilon_c x_\rho}.
\end{equation}
Thus the dipole-dipole transfer amplitude scales as
\begin{equation}
 \mathcal V_{DA}^{\mathrm{ON}}\propto x_\rho^{-3},
\end{equation}
while the transfer rate scales as
\begin{equation}
 \Gamma_{D\to A}^{\mathrm{ON}}\propto x_\rho^{-6}.
\label{eq:FRET_ON_power}
\end{equation}

\subsection{Dirichlet/PEC branch: exponential FRET screening}

For $r_{\TM}\to-1$, Eq.~\eqref{eq:S_minus_Bessel} gives an odd transverse spectrum.  The lowest transverse harmonic controls the large-distance envelope.  For $x_\rho\gg W$,
\begin{align}
D_{\mathrm{OFF}}^{(-)}(x_\rho)
&\simeq \frac{1}{2\pi\varepsilon_0\varepsilon_c W}K_0(\pi x_\rho/W)
\nn\\
&\sim \frac{1}{2\pi\varepsilon_0\varepsilon_c W}\sqrt{\frac{W}{2x_\rho}}e^{-\pi x_\rho/W}.
\end{align}
After differentiation, the dipole amplitude contains $K_1$ and/or $K_2$ depending on orientation, but all components share the same exponential envelope,
\begin{equation}
 \mathcal V_{DA}^{(-)}(x_\rho)
 \propto e^{-\pi x_\rho/W}.
\end{equation}
Therefore
\begin{equation}
 \Gamma_{D\to A}^{(-)}(x_\rho)
 \propto e^{-2\pi x_\rho/W}.
\label{eq:FRET_screened_exp}
\end{equation}
This is the direct FRET analogue of exponential screening of static Coulomb and dipole interactions.  The decay length is set by the spacer thickness,
\begin{equation}
 \ell_{\mathrm{scr}}=\frac{W}{\pi},
\end{equation}
for the lowest odd mode.

\subsection{Neumann/PMC-like branch: logarithmic antiscreening}

For $r_{\TM}=1-\delta$ with $0<\delta\ll1$, the scalar Green function enters the quasi-two-dimensional regime in Eq.~\eqref{eq:log_antiscreening}.  Differentiating the logarithm gives a dyadic amplitude scaling as
\begin{equation}
 \mathcal V_{DA}^{(+)}(x_\rho)
 \sim \frac{d_A d_D}{2\pi\varepsilon_0\varepsilon_c W x_\rho^2},
 \qquad
 W\ll x_\rho\ll x_{\rho,*}.
\label{eq:VDA_log_dyadic}
\end{equation}
The corresponding FRET rate scales as
\begin{equation}
 \Gamma_{D\to A}^{(+)}(x_\rho)
 \propto \frac{1}{W^2 x_\rho^4}.
\label{eq:FRET_log_scaling}
\end{equation}
Relative to the bulk near-field law $x_\rho^{-6}$, the quasi-two-dimensional branch enhances the rate by the approximate factor
\begin{equation}
 \frac{\Gamma_{D\to A}^{(+)}}{\Gamma_{D\to A}^{\mathrm{ON}}}
 \sim \left(\frac{x_\rho}{W}\right)^2,
 \qquad
 W\ll x_\rho\ll x_{\rho,*}.
\label{eq:FRET_enhancement_factor}
\end{equation}
The usable range is gate-programmable:
\begin{equation}
 x_{\rho,*}\sim\frac{W}{1-r_{\TM}}.
\end{equation}
This is the FRET counterpart of logarithmic antiscreening: a near-gapless transverse mode extends the effective dimensionality of the image lattice and converts the $1/x_\rho^3$ dipole amplitude into a quasi-2D $1/(W x_\rho^2)$ amplitude.

\begin{table*}[t]
\caption{Universal near-field tuning regimes for the donor-acceptor coupling and the FRET rate.  Here $\eta=x_\rho/W$.  The index $\nu$ in the Bessel functions depends on dipole orientation; the exponential envelope is universal.}
\begin{ruledtabular}
\begin{tabular}{lll}
Regime & Donor-acceptor amplitude $\mathcal V_{DA}$ & FRET rate $\Gamma_{D\to A}$ \\
\hline
Transparent, $r_{\TM}\to0$ & $x_\rho^{-3}$ & $x_\rho^{-6}$ \\
Dirichlet/PEC, $r_{\TM}\to-1$ & $K_{\nu}(\pi x_\rho/W)\sim \eta^{-1/2}e^{-\pi\eta}$ & $K_{\nu}^2(\pi x_\rho/W)\sim \eta^{-1}e^{-2\pi\eta}$ \\
Neumann/PMC-like, $r_{\TM}\to1^-$ & $(W x_\rho^2)^{-1}$ over $W\ll x_\rho\ll x_{\rho,*}$ & $(W^2 x_\rho^4)^{-1}$ over $W\ll x_\rho\ll x_{\rho,*}$ \\
\end{tabular}
\end{ruledtabular}
\label{tab:FRET_regimes}
\end{table*}

Table~\ref{tab:FRET_regimes} summarizes the three universal near-field regimes and makes explicit how the change in scalar Green-function dimensionality translates into the donor--acceptor amplitude and the corresponding FRET-rate law.

\section{Applications and experimental relevance}
\label{sec:applications}

\subsection{Moir\'e systems}
The formalism is especially natural for excitonic, molecular, and
rare-earth emitters embedded in van der Waals heterostructures.  Interlayer
excitons in TMD bilayers carry out-of-plane electric dipoles and therefore
couple strongly to the TM sector, which is the sector most strongly
controlled by gate-tunable charge response.  Recent experiments have
established strongly correlated excitonic insulators and exciton crystals
in TMD double-layer and moir\'e platforms \cite{Ma2021,Qi2026}, while
theory has explored collective exciton transport and hydrodynamic regimes
in such materials \cite{Mantsevich2024}.  The present architecture adds a
new electromagnetic control knob: the transfer of excitation between
localized excitons, moir\'e traps, molecular dopants, or rare-earth ions can
be switched from bulk-like \(x_\rho^{-6}\) behavior to plasmon-assisted
enhancement or exponential suppression without changing the donor or
acceptor themselves.

\subsection{Spontaneous emission tuning using graphene}
A first experimentally accessible consequence is the gate-tunable
spontaneous-emission rate of a single emitter.  The spontaneous-emission
rate is a local quantity governed by the imaginary part of the retarded
Green tensor at the emitter position,
\begin{equation}
F_P(\bx;E_F)
=
\frac{\Gamma(\bx;E_F)}{\Gamma_{\rm bulk}}
=
\frac{
\bd^{\,*}\cdot
\Im\obG_{\rm R}(\bx,\bx;\omega_0;E_F)
\cdot
\bd
}{
\bd^{\,*}\cdot
\Im\obG_{\rm bulk}(\bx,\bx;\omega_0)
\cdot
\bd
}.
\label{eq:Purcell_factor_applications}
\end{equation}
This local Green-tensor channel has already been observed experimentally in an Er-graphene platform: Cano \textit{et al.} measured strong gate-dependent Er decay into graphene and showed that the microscopic decay pathway can be switched between interband electron-hole excitations and intraband/plasmonic channels by tuning $E_F$ \cite{Cano2020ErGraphene}.  Although their geometry contained a single graphene sheet rather than the double-layer cavity considered here, their result directly validates the strong, electrically tunable Er-graphene near-field coupling on which the present implementation builds.

For the graphene-dielectric-graphene structure, the separation between
the two graphene sheets is the dielectric spacer thickness \(W\), with the
sheets located at \(x_z=\pm W/2\).  For an emitter at coordinate \(x_z\),
the distances to the upper and lower graphene sheets are
\(W/2-x_z\) and \(W/2+x_z\), respectively, and the distance to the nearest
sheet is \(W/2-|x_z|\).  These vertical distances should not be confused
with the lateral donor-acceptor separation \(x_\rho\).  At the midplane,
\(x_z=0\), the emitter is a distance \(W/2\) from each graphene sheet.

\begin{figure}[htb] 
\centering 
\includegraphics[width=0.50\textwidth]{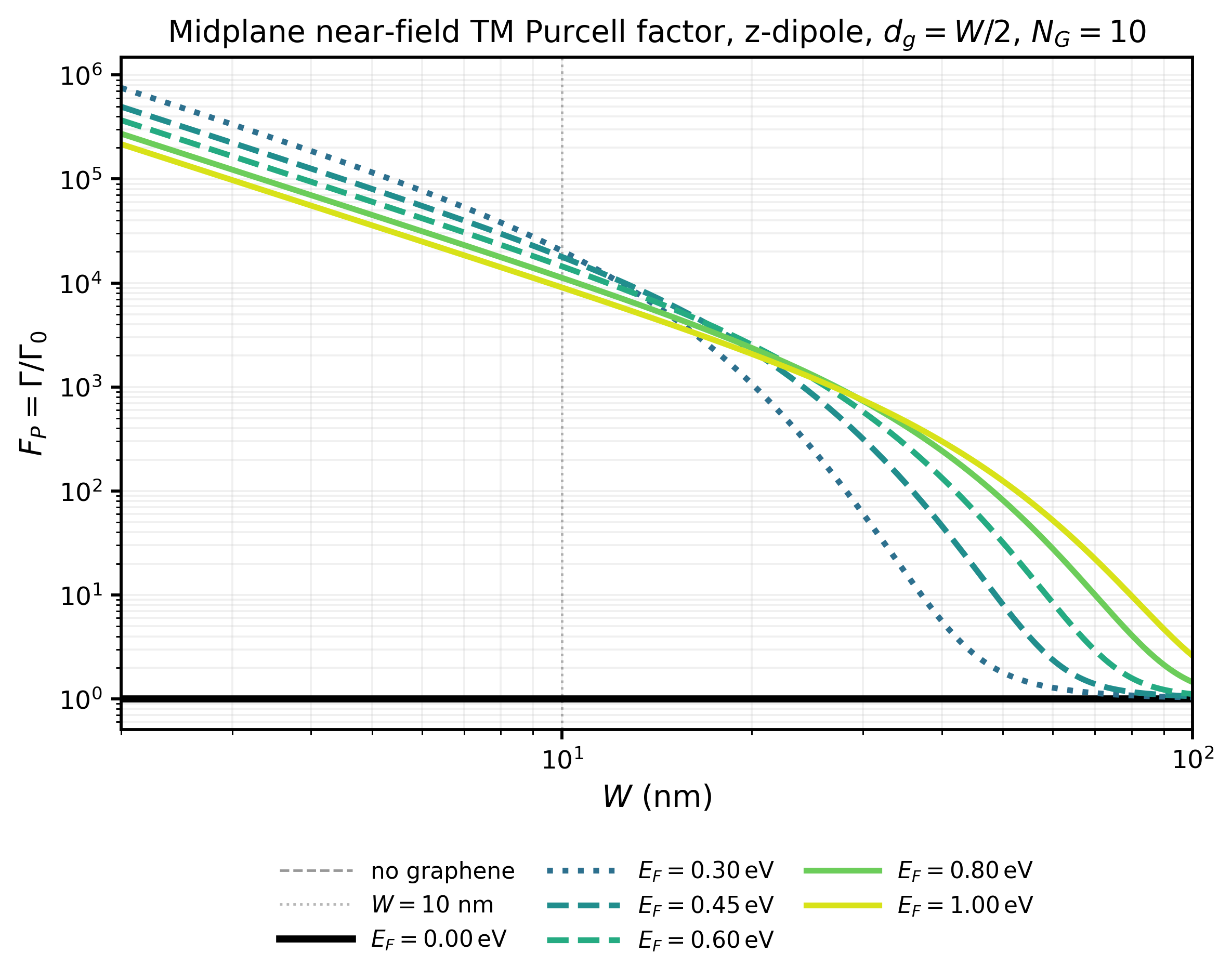} 
\caption{ Gate-tunable spontaneous emission of a single Er emitter at the center of a double multilayer-graphene cavity. The normalized spontaneous-emission rate, or Purcell factor, \(F_P=\Gamma/\Gamma_{\rm bulk}\), is plotted as a function of the dielectric spacer thickness \(W\), which is also the graphene-graphene separation. The emitter is fixed at the midplane, so its distance to each graphene mirror is \(W/2\). Each mirror is modeled as an effective multilayer graphene sheet with \(N_G=10\), corresponding to \(\sigma_{\rm eff}=N_G\sigma_g\). At small \(W\), the emitter near field overlaps strongly with evanescent TM channels of both graphene mirrors, leading to a large, gate-tunable modification of the local density of optical states. Increasing \(W\) reduces this near-field overlap and drives the emission rate back toward the homogeneous bulk value. Electrostatic tuning of the graphene Fermi energy \(E_F\) therefore controls the Er lifetime at fixed emitter position by changing the graphene TM response at the Er transition frequency. } 
\label{fig:Er_Purcell_W_midplane} 
\end{figure}

Figure~\ref{fig:Er_Purcell_W_midplane} shows the normalized spontaneous-emission rate of a single Er emitter placed at the center of the graphene-dielectric-graphene cavity as a function of the spacer thickness \(W\). In this geometry the emitter remains equidistant from the two graphene mirrors, with a distance \(W/2\) to each sheet, so changing \(W\) directly changes the evanescent near-field overlap between the emitter and the two conducting boundaries. For small \(W\), the emitter couples efficiently to high-\(q_\rho\) TM channels of both multilayer graphene mirrors, producing a large Purcell enhancement whose magnitude is controlled by the graphene Fermi energy \(E_F\). As \(W\) is increased, the emitter is moved farther from both sheets, the evanescent overlap decreases, and \(F_P\) approaches the homogeneous bulk value. This midplane geometry is the same geometry used for the Er-Er FRET calculation below; therefore Fig.~\ref{fig:Er_Purcell_W_midplane} directly quantifies the local single-emitter decay environment that accompanies the nonlocal FRET interaction.

\subsection{FRET tuning using graphene}
The same retarded Green tensor also controls the nonlocal FRET channel
between two emitters.  For two Er ions placed in the midplane,
\(x_z=x_z'=0\), the local emitter-graphene distance is fixed at
\(W/2\), while the FRET rate depends on the lateral donor-acceptor
separation \(x_\rho\):
\begin{equation}
\Gamma_{D\to A}(x_\rho;E_F)
\propto
\left|
\bd_A^{\,*}\cdot
\obG_{\rm R}(\bx_A,\bx_D;\omega_D;E_F)
\cdot
\bd_D
\right|^2 .
\label{eq:FRET_applications_nonlocal}
\end{equation}
This distinction is important experimentally.  Spontaneous emission probes
the local spectral density \(\Im\obG_{\rm R}(\bx,\bx;\omega_0)\), whereas
FRET probes the nonlocal propagation amplitude
\(\obG_{\rm R}(\bx_A,\bx_D;\omega_D)\).  Both quantities are controlled by
the same graphene reflection amplitudes \(r_{\rm TM/TE}(q_\rho,\omega;E_F)\),
but they sample different spatial arguments of the same Green tensor.

The measurements of Cano \textit{et al.} establish experimentally the local Er$\rightarrow$graphene part of this sequence \cite{Cano2020ErGraphene}.  The present two-emitter problem adds a nonlocal second step: the donor-generated graphene-assisted field propagates to the acceptor.  If real graphene excitations are available at $\omega_{\rm Er}$, this channel contains a dissipative on-shell contribution associated with electron-hole excitations or plasmon launching.  If the graphene-plasmon states are detuned or gapped, the graphene contribution becomes predominantly dispersive and virtual, allowing the nonlocal Er-Er coupling to remain finite while suppressing direct Er-to-graphene loss.  This distinction is particularly important when graphene is used not merely as an absorber, but as a mediator of emitter-emitter interactions.

For a single graphene sheet, the Er-Er FRET modulation at
\(\lambda_{\rm Er}\simeq1.53~\mu{\rm m}\) is modest for realistic Fermi
energies, because the near-field TM reflection amplitude does not yet reach
the Dirichlet-like limit \(r_\phi\to -1\) over the momentum window sampled
by the FRET Green function.  This motivates using multilayer or
intercalated graphene mirrors.  In the thin-multilayer limit, where the
total stack thickness is small compared with the relevant evanescent decay
length, the in-plane sheet currents add in parallel.  The most general
additive-sheet form is therefore
\begin{equation}
\sigma_{\rm eff}(q_\rho,\omega)
=
\sum_{\ell=1}^{N_G}
\sigma_{\ell}(q_\rho,\omega),
\label{eq:applications_sigma_eff_multilayer_general}
\end{equation}
where \(\sigma_{\ell}\) is the conductivity of graphene layer \(\ell\).
When \(N_G\) electronically active layers have approximately the same
carrier density, scattering rate, and optical conductivity, Eq.~\eqref{eq:applications_sigma_eff_multilayer_general}
reduces to
\begin{equation}
\sigma_{\rm eff}(q_\rho,\omega;E_F)
\simeq
N_G\,\sigma_g(q_\rho,\omega;E_F).
\label{eq:applications_sigma_eff_multilayer}
\end{equation}
Thus \(N_G\) should be interpreted as the effective number of equivalent
conducting graphene layers rather than, in every material, the literal
number of carbon sheets.

Several experimentally established multilayer graphene systems motivate
this approximation.  Multilayer epitaxial graphene grown on the
carbon-terminated face of SiC develops rotational stacking faults that
strongly suppress Bernal-type interlayer hybridization, so that films tens
of layers thick retain monolayer-like Dirac electronic structure and very
high mobility \cite{deHeer2007EpitaxialGraphene,Hass2008MultilayerGraphene,Orlita2008HighMobilityMEG}.  FeCl$_3$ intercalation provides an even more direct route to electronically decoupled, strongly hole-doped graphene sheets: Raman and band-structure studies show that adjacent graphene layers recover nearly monolayer-like dispersion after intercalation \cite{Zhan2010FeCl3FLG,Zhao2011FeCl3Intercalation}, while transport measurements reveal multiple parallel two-dimensional hole gases and exceptionally low sheet resistance \cite{Khrapach2012FeCl3Conductors,Bointon2015FeCl3CVD}.  In our earlier ACS Nano work on nanopatterned FeCl$_3$-intercalated multilayer graphene, we used this additive multilayer response to obtain an approximately \(N_G\)-proportional unpatterned optical response and a strongly enhanced tunable plasmonic response \cite{Shabbir2022FeCl3Multilayer}.

A related and newer route uses the large work-function mismatch between
graphene and \(\alpha\)-RuCl$_3$.  Graphene/\(\alpha\)-RuCl$_3$ interfaces exhibit massive charge-transfer hole doping of graphene and support charge-transfer plasmon polaritons with \(E_F\) of order \(0.6~\mathrm{eV}\) \cite{Rizzo2020RuCl3Plasmons}.  Very recently, \(\alpha\)-RuCl$_3$ has also been intercalated directly into graphite: stage-2 and stage-4 compounds have been synthesized, and transport together with first-principles calculations shows substantial charge transfer with comparatively weak graphene--RuCl$_3$ hybridization \cite{Razpopov2026RuCl3Intercalated}.  This makes RuCl$_3$ intercalation a promising additional route to high-carrier-density multilayer graphene mirrors.  However, a simple \(N_G\sigma_g\) scaling has not yet been established experimentally for the RuCl$_3$ intercalate; for that material Eq.~\eqref{eq:applications_sigma_eff_multilayer_general} is the appropriate starting point.
Supplementary Information Sec.~S12 places these graphene implementations in the broader context of electrostatically gated conductors, phase-change and correlated-oxide mirrors, transparent conducting oxides, and tunable high-impedance magnetic mirrors.

In the equivalent-layer limit of Eq.~\eqref{eq:applications_sigma_eff_multilayer}, the Drude weight increases approximately by \(N_G\) and the graphene plasmon pole shifts to smaller wave vector,
\begin{equation}
q_{\rm pl}^{(N_G)}(\omega;E_F)
\simeq
\frac{1}{N_G}\,
q_{\rm pl}^{(1)}(\omega;E_F).
\label{eq:applications_qpl_multilayer}
\end{equation}
Consequently, the condition for reflective near-field screening,
\begin{equation}
q_\ast\gg q_{\rm pl}^{(N_G)}(\omega_{\rm Er};E_F),
\qquad
q_\ast\simeq \frac{\pi}{W},
\label{eq:applications_reflective_condition_NG}
\end{equation}
can be reached at experimentally reasonable Fermi energies by increasing
\(N_G\).  This is the two-dimensional analogue of making the Er transition
frequency lie below the plasma frequency of the mirror at the spatial
Fourier component relevant to FRET.

\begin{figure}[htb]
\centering
\includegraphics[width=0.50\textwidth]{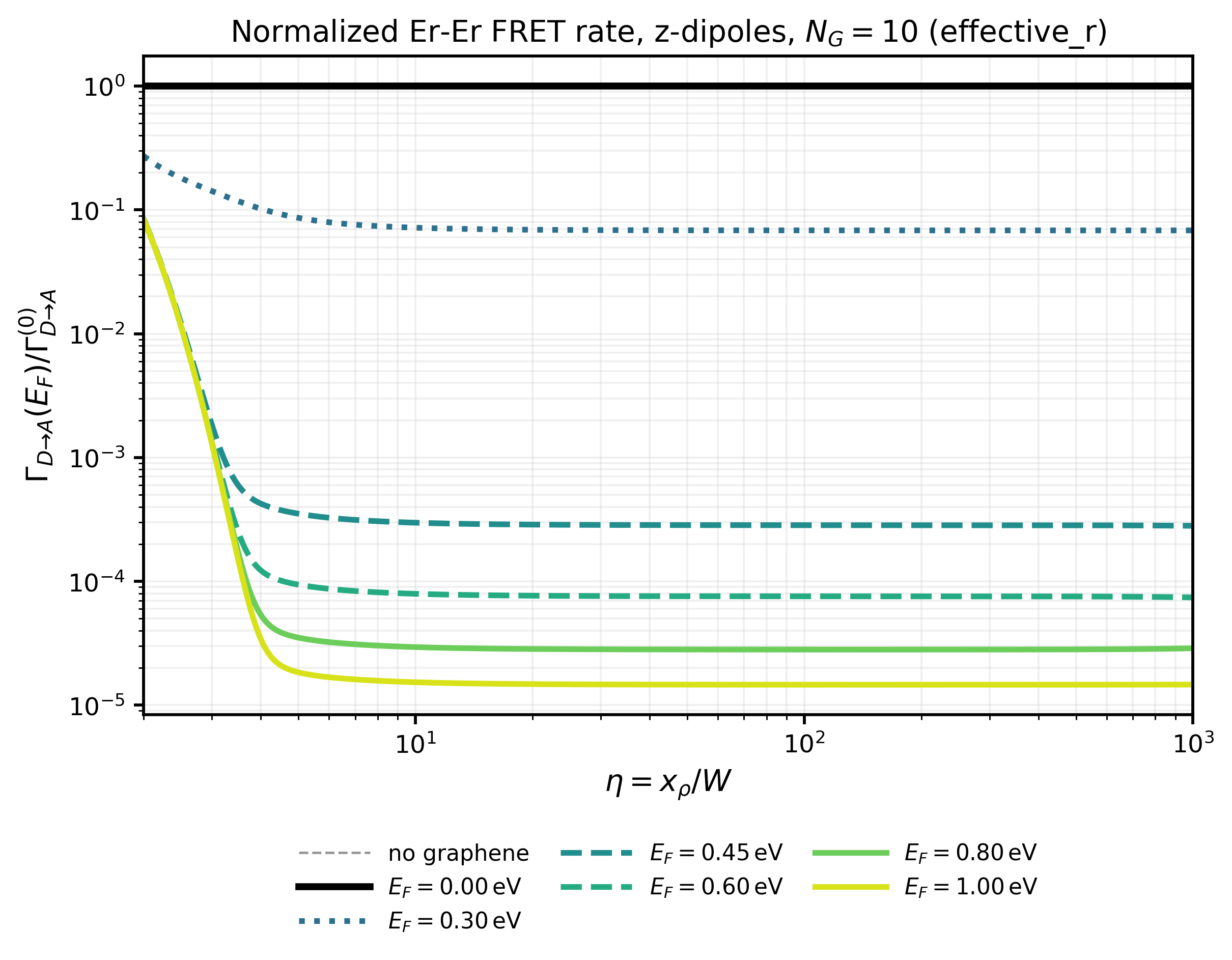}
\caption{
Gate-tunable Er-Er FRET in a double multilayer-graphene cavity.  The
normalized FRET rate
\(\Gamma_{D\to A}(E_F)/\Gamma_{D\to A}^{(0)}\) is plotted as a function of
the normalized lateral separation \(\eta=x_\rho/W\) for two Er emitters in
the midplane.  Each mirror is modeled as an effective multilayer graphene
sheet with \(N_G=10\), so that
\(\sigma_{\rm eff}=N_G\sigma_g\).  Increasing \(E_F\) increases the
reactive Drude weight, moves the graphene plasmon pole to smaller
\(q_\rho\), and drives the scalar TM reflection amplitude toward the
Dirichlet-like limit \(r_\phi\to-1\).  In this reflective branch the
midplane Green function acquires the screened Bessel-\(K\) envelope, leading
to strong suppression of FRET at \(\eta\gtrsim 2\).  The plotted range is
restricted to \(2<\eta<10^3\), where the long-distance screening model is
most reliable.
}
\label{fig:Er_FRET_NG10}
\end{figure}

Figure~\ref{fig:Er_FRET_NG10} shows the resulting nonlocal FRET tuning for
\(N_G=10\).  In this regime, increasing \(E_F\) no longer merely increases
the plasmonic density of states.  Instead, the enhanced multilayer Drude
weight drives the TM scalar reflection amplitude toward
\(r_\phi\simeq -1\), producing the reflective Dirichlet-like branch.  The
midplane scalar Green function then crosses over from the transparent
near-field power law to the screened Bessel-\(K\) form,
\begin{equation}
\calG_\phi^{D}(x_\rho)
\propto
K_0\!\left(\pi\frac{x_\rho}{W}\right)
\sim
\frac{
e^{-\pi x_\rho/W}
}{
\sqrt{x_\rho/W}
},
\qquad
x_\rho\gg W.
\label{eq:applications_Dirichlet_screened_backbone}
\end{equation}
After the electric-field derivatives entering the dipole Green tensor are
applied, the normalized Er-Er FRET rate inherits an approximately squared
exponential envelope,
\begin{equation}
\frac{
\Gamma_{D\to A}^{D}
}{
\Gamma_{D\to A}^{(0)}
}
\sim
\eta^6
\left[
K_0(\pi\eta)
\right]^2
\sim
\frac{\eta^5}{2}
e^{-2\pi\eta},
\qquad
\eta=\frac{x_\rho}{W}\gg1,
\label{eq:applications_FRET_screening_asymptotic}
\end{equation}
up to orientation-dependent prefactors.  The \(N_G=10\) result therefore demonstrates a practically accessible route to suppressing Er-Er FRET by as much as \(10^5\) using electrostatically tunable graphene-based mirrors.
The analytical Er--Er Green-function and rate formulas underlying this calculation are derived in Supplementary Information Sec.~S9, and the numerical evaluation procedure is given in Sec.~S10 and Appendix~\ref{app:numerics}.

Measuring the Purcell factor and the FRET rate in the same device would
therefore provide a stringent test of the theory.  The single-emitter
measurement determines how the graphene sheets tune the local density of
states and the Er lifetime as a function of \(W\), the emitter position \(x_z\), and \(E_F\).  The
two-emitter measurement then tests the nonlocal prediction that the same
gate-tunable electromagnetic environment can reshape the donor-acceptor
transfer law from transparent bulk-like behavior to a plasmon-assisted
regime or to the reflective Dirichlet-like branch with exponential
screening.  In this way, graphene-dielectric-graphene heterostructures offer a
unified platform for controlling both local emission and nonlocal energy
transfer with a single electrostatic knob.  Rotationally decoupled C-face
epitaxial graphene, FeCl$_3$-intercalated multilayer graphene, and emerging
RuCl$_3$-intercalated graphite provide complementary materials routes for
realizing the high effective sheet conductivities required by the
multilayer-mirror regime discussed above.

\begin{figure*}[htb]
\centering
\includegraphics[width=1.0\textwidth]{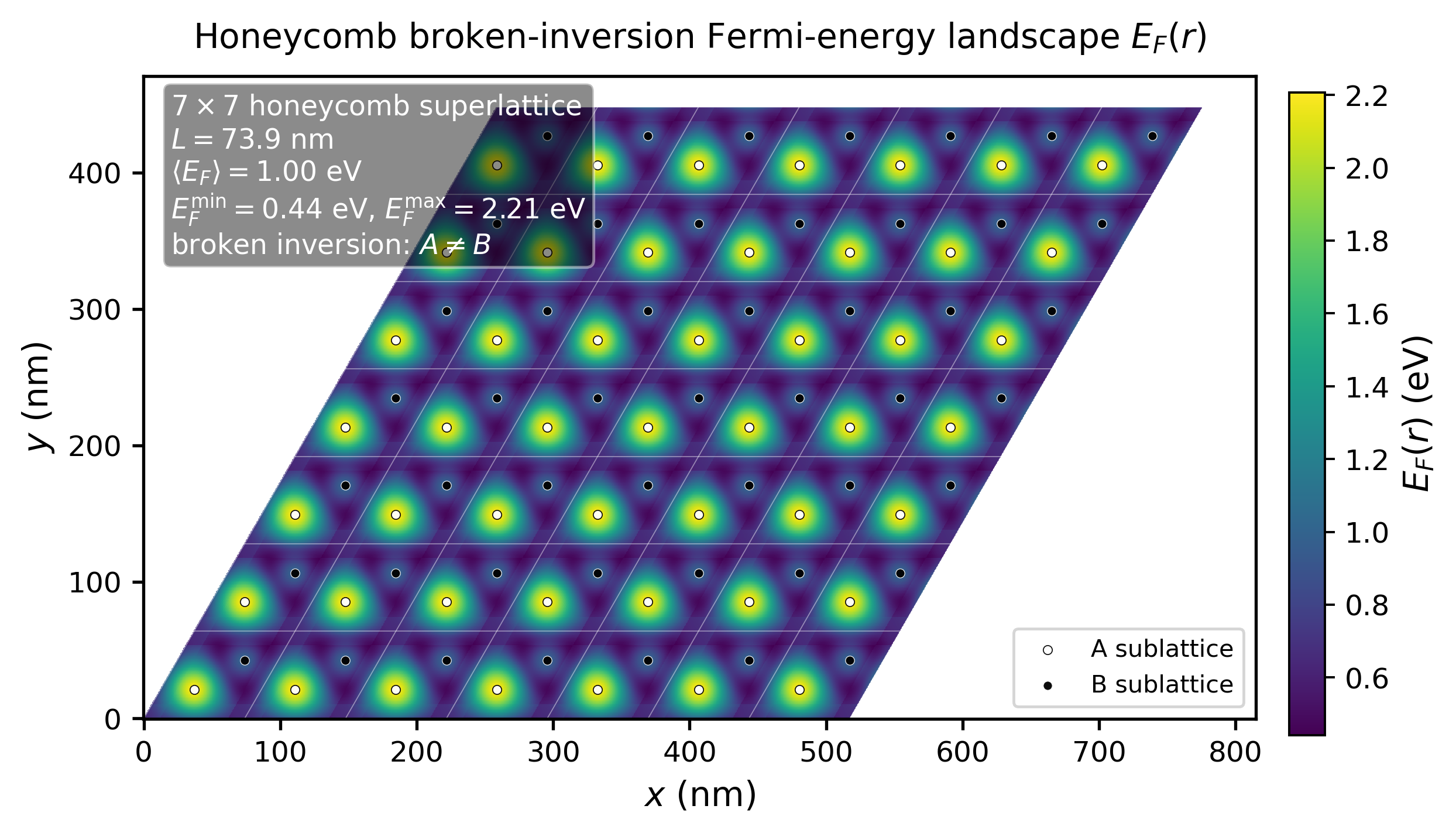}
\caption{ \textbf{Honeycomb broken-inversion Fermi-energy landscape.} Real-space \(7\times7\) superlattice of the metagate-induced graphene Fermi energy \(E_F(\mathbf r)\) used for the plasmonic band-structure calculation in Fig.~\ref{fig:SI_Er_DLG_honeycomb_bandstructure}. The pattern is an hBN-like honeycomb lattice with two inequivalent sublattices, \(A\) and \(B\), which break inversion symmetry and open a mass gap in the plasmonic Dirac spectrum at \(K\) and \(K'\). The color scale shows the local Fermi energy, while the white and black markers indicate the two sublattice centers. The large average carrier density can be supplied by an ion-gel or electrolyte back gate, while the nearby patterned metagate creates the nanoscale honeycomb modulation. This architecture is designed to place the Er transition inside a bulk plasmonic band gap, with no propagating graphene-plasmon Bloch mode at \(\omega_{\rm Er}\). } \label{fig:SI_Er_DLG_honeycomb_EF_superlattice}
\end{figure*}

\subsection{Dissipation-suppressed Er-Er FRET in a metagate-induced graphene-plasmon band gap}
\label{subsec:Er_FRET_metagate_gap}

The preceding FRET-suppression mechanism assumes that the graphene
mirrors act predominantly as reactive electromagnetic boundaries at the
Er transition frequency.  A potential loss channel remains, however, if
the Er transition is resonant with a propagating graphene-plasmon mode.  This
is not merely a theoretical concern: Cano \textit{et al.} directly observed
strong gate-controlled Er-to-graphene energy transfer and, above the
Pauli-blocking threshold, identified intraband/plasmon generation as the
dominant Er decay pathway \cite{Cano2020ErGraphene}.  Their experiment
therefore provides a direct benchmark for the dissipative on-shell channel
that the band-gap design below is intended to suppress.
The corresponding plasmonic-crystal Green-function construction is detailed in Appendix~\ref{app:metagate_tuned_DLG_green_functions}: Appendix~\ref{app:mode_resolved_green_function} gives the mode-resolved spectral sum, Appendix~\ref{app:directional_green_cuts} defines the directional real-space cuts, and Appendix~\ref{app:green_decay_results_reduced} compares in-gap evanescence with band-edge propagation.
For translationally invariant graphene, the symmetric and antisymmetric
double-layer plasmon branches (often labeled as optical and acoustic branches \cite{GoncalvesPeres2016}) are gapless, so a finite transition
frequency \(\omega_{\rm Er}\) generically intersects the plasmon
dispersion at one or more real in-plane momenta.  These intersections are
on-shell plasmon channels and can produce plasmon-assisted energy transfer
or nonradiative loss.

To remove these on-shell channels, we replace the continuous
graphene-plasmon spectrum by a plasmonic-crystal spectrum with a complete
band gap at \(\omega_{\rm Er}\).  Following the metagate concept of
Jung, Fan, and Shvets~\cite{PhysRevLett.121.086807}, a patterned metallic
gate placed a few nanometers from graphene can imprint a periodic
Fermi-energy landscape \(E_F(\brho)\) without physically etching the
graphene.  Since the graphene Drude weight is proportional to
\(E_F(\brho)\), this produces a periodic plasmonic refractive-index
landscape and opens Bragg gaps in the graphene-plasmon bands.  Here we use
the same principle for a different purpose: instead of creating
topological domain-wall plasmon waveguides, we use a uniform
band-gap domain to eliminate propagating plasmon modes at the Er
transition.

\begin{figure*}[htb]
\centering
\includegraphics[width=1.0\textwidth]{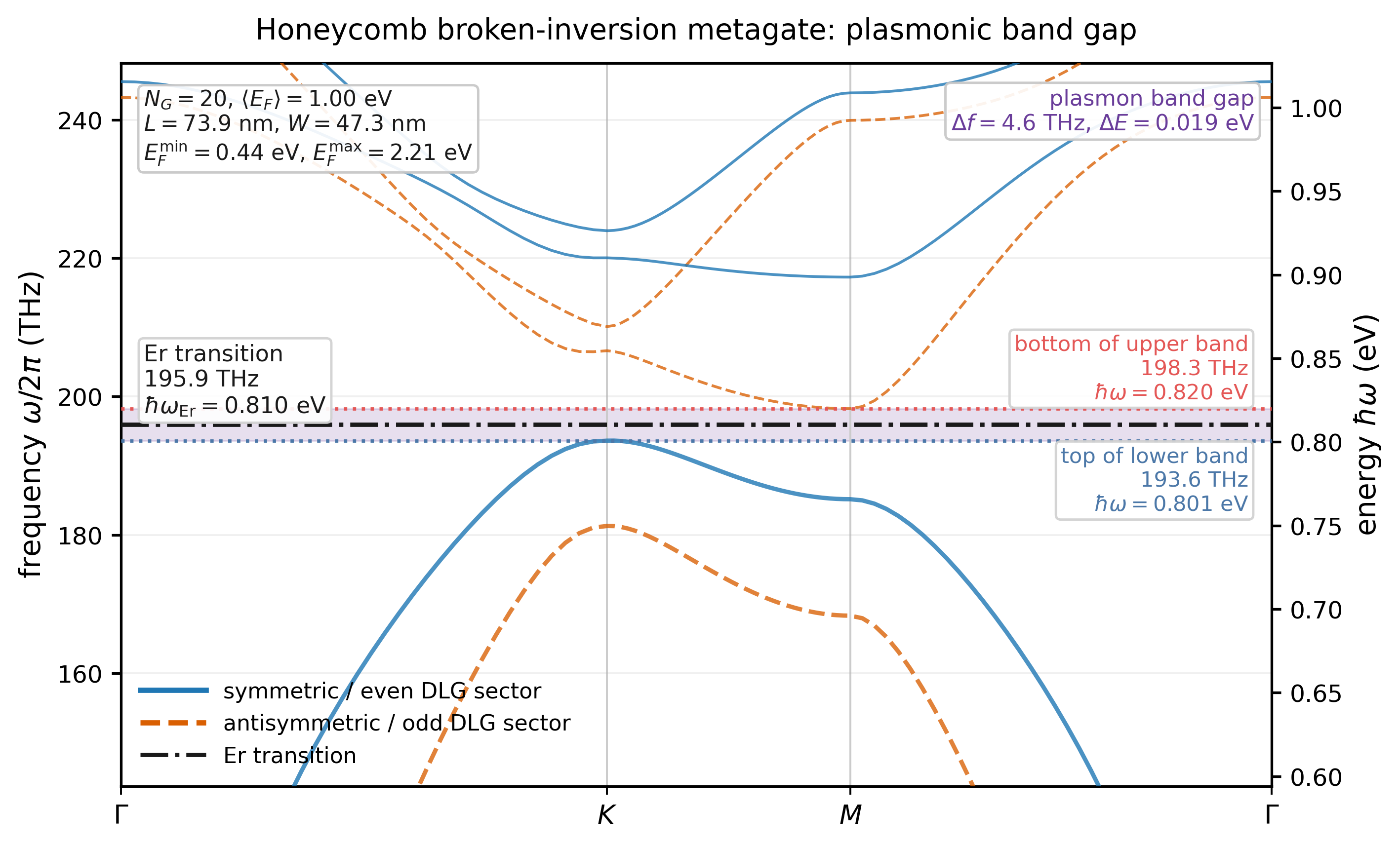}
\caption{ \textbf{Metagate-induced plasmon band gap at the Er transition.} Calculated plasmonic-crystal band structure of an effective double-layer multilayer-graphene mirror with a honeycomb broken-inversion metagate. The antisymmetric, odd layer sector and symmetric, even layer sector are shown along the \(\Gamma-K-M-\Gamma\) path of the triangular Brillouin zone. The metagate imposes an hBN-like honeycomb Fermi-energy landscape with inequivalent \(A\) and \(B\) sublattices, opening a plasmonic band gap at the valley points. The horizontal dash-dotted line marks the Er transition frequency \(\omega_{\rm Er}/2\pi\), corresponding to \(\lambda_{\rm Er}=1530~{\rm nm}\) or \(\hbar\omega_{\rm Er}\simeq0.81~{\rm eV}\). The dotted horizontal lines mark the top of the lower plasmon band, or plasmonic valence band, and the bottom of the upper plasmon band, or plasmonic conduction band. The shaded region is the complete plasmon band gap used to suppress on-shell graphene-plasmon loss channels. Since the Er transition lies inside this gap, Er-Er FRET remains resonant between donor and acceptor, while the graphene plasmons contribute only virtually through the reactive part of the retarded Green tensor. } \label{fig:SI_Er_DLG_honeycomb_bandstructure}
\end{figure*}

The specific design considered here is an hBN-like honeycomb metagate with
broken inversion symmetry.  The honeycomb unit cell contains two
inequivalent sublattices, \(A\) and \(B\), which impose different local
Fermi-energy environments on the graphene,
\[
E_F^{A}\neq E_F^{B}.
\]
This is the plasmonic analogue of inversion-symmetry breaking in hBN:
when the two sublattices are equivalent, the honeycomb plasmonic crystal
has Dirac-like degeneracies at \(K\) and \(K'\); when the two sublattices
are inequivalent, a mass term opens a gap.  A convenient smooth
parametrization of the Fermi-energy landscape is
\begin{equation}
\begin{split}
E_F(\brho) = {}& \overline E_F + \Delta_F \frac{f_A(\brho)-f_B(\brho)}{f_A(\brho)+f_B(\brho)+s_0} \\
& + E_c \left[ f_A(\brho)+f_B(\brho) - \overline{f_A+f_B} \right],
\end{split}
\label{eq:honeycomb_EF_landscape}
\end{equation}
where
\begin{equation}
f_{\nu}(\brho)
=
\sum_{\mathbf R}
\exp\left[
-\frac{
|\brho-\mathbf R-\boldsymbol\tau_\nu|^2
}{
2\sigma^2
}
\right],
\qquad
\nu=A,B .
\label{eq:honeycomb_gaussian_sublattice}
\end{equation}
Here \(\mathbf R\) runs over the triangular Bravais lattice of the
honeycomb superlattice, \(\boldsymbol\tau_A\) and
\(\boldsymbol\tau_B\) are the two basis positions, \(\Delta_F\) controls
the inversion-breaking contrast, and \(E_c\) controls a common honeycomb
modulation.  Figure~\ref{fig:SI_Er_DLG_honeycomb_EF_superlattice} shows
the resulting \(7\times7\) superlattice used in the numerical example.

We model each graphene mirror as an effective multilayer graphene sheet
with \(N_G\) electronically active layers.  In the local Drude
approximation this amounts to multiplying the Drude weight by \(N_G\).
For a plasmon Bloch wave vector \(\bk\), the potential on either graphene
sheet is expanded as
\begin{equation}
\Phi(\brho)
=
\sum_\alpha
\phi_{\alpha}
e^{\I(\bk+\mathbf G_\alpha)\cdot\brho},
\label{eq:honeycomb_phi_Bloch}
\end{equation}
where \(\mathbf G_\alpha\) are reciprocal lattice vectors.  The
Fermi-energy modulation enters through the density-response matrix
\begin{equation}
\left[
\mathbb X_{\bk}
\right]_{\alpha\beta}
=
\frac{N_G}{\pi\hbar^2}
(\bk+\mathbf G_\alpha)\cdot(\bk+\mathbf G_\beta)
\widetilde E_F(\mathbf G_\alpha-\mathbf G_\beta),
\label{eq:honeycomb_X_matrix}
\end{equation}
with
\begin{equation}
\widetilde E_F(\mathbf G)
=
\frac{1}{\Omega}
\int_{\Omega}
d^2\rho\,
E_F(\brho)e^{-\I\mathbf G\cdot\brho}.
\label{eq:honeycomb_EF_fourier}
\end{equation}
For a symmetric double-layer graphene structure with the two sheets located at
\(x_z=\pm W/2\), the symmetric and antisymmetric sectors are obtained from the even and
odd layer combinations.  In a homogeneous dielectric approximation the
corresponding Coulomb propagators are
\begin{equation}
\left[
\mathbb V_{\bk}^{(\pm)}
\right]_{\alpha\beta}
=
\delta_{\alpha\beta}
\frac{e^2}{2\varepsilon_0\varepsilon_{\rm eff}q_\alpha}
\left[
1\pm e^{-q_\alpha W}
\right],
\label{eq:honeycomb_DLG_Coulomb_pm}
\end{equation}
where $q_\alpha=|\bk+\mathbf G_\alpha|$.
The \(+\) sector is the symmetric double-layer plasmon branch,
whereas the \(-\) sector is the antisymmetric plasmon branch.  The
plasmonic-crystal bands are then obtained from
\begin{equation}
\mathbb V_{\bk}^{(\pm)}
\mathbb X_{\bk}
|\phi_{n\bk}^{(\pm)}\rangle
=
\omega_{n\bk,\pm}^2
|\phi_{n\bk}^{(\pm)}\rangle .
\label{eq:honeycomb_DLG_eigenproblem}
\end{equation}
In the numerical calculation this generalized eigenproblem is solved along
the \(\Gamma-K-M-\Gamma\) path of the triangular Brillouin zone.

Figure~\ref{fig:SI_Er_DLG_honeycomb_bandstructure} shows the resulting
plasmonic band structure for an effective \(N_G=20\) multilayer graphene
mirror.  The Er transition line lies inside the inversion-breaking plasmon
band gap, between the top of the lower plasmon band and the bottom of the
upper plasmon band.  The operating condition is therefore
\begin{equation}
\omega_v+\Delta_\gamma
<
\omega_{\rm Er}
<
\omega_c-\Delta_\gamma,
\label{eq:Er_inside_honeycomb_gap}
\end{equation}
where \(\omega_v\) is the top of the lower band, \(\omega_c\) is the
bottom of the upper band, and \(\Delta_\gamma\sim\gamma/2\) is a damping
margin.  Equivalently,
\begin{equation}
\omega_{n\bk,\pm}\neq\omega_{\rm Er}
\qquad
\text{for all }n,\bk .
\label{eq:no_on_shell_honeycomb_plasmons}
\end{equation}
Thus there is no propagating graphene-plasmon Bloch mode available at the
Er transition frequency.  This operating point is complementary to the
plasmon-launching regime demonstrated experimentally by Cano \textit{et al.}
\cite{Cano2020ErGraphene}: rather than resonantly converting Er excitation
into a real propagating graphene plasmon, the present design uses the same
Er-graphene interaction off shell, so that graphene contributes primarily
through virtual, reactive plasmonic states.

Inside the plasmonic band gap, the graphene-plasmon contribution to the
retarded Green tensor is off resonant.  In spectral form,
\begin{equation}
\obG_R^{\rm pl}(\omega)
=
\sum_{n,\bk,\lambda=\pm}
\frac{
\obF_{n\bk,\lambda}
}{
\omega_{n\bk,\lambda}^2
-
(\omega+\I0^+)^2
-
\I\gamma_{n\bk,\lambda}\omega
}.
\label{eq:honeycomb_plasmon_spectral_G}
\end{equation}
When Eq.~\eqref{eq:no_on_shell_honeycomb_plasmons} is satisfied, no
denominator vanishes at \(\omega=\omega_{\rm Er}\).  Consequently,
\begin{equation}
\Im\obG_R^{\rm pl}(\omega_{\rm Er})
\sim
\sum_{n,\bk,\lambda}
\frac{
\gamma_{n\bk,\lambda}\omega_{\rm Er}\,
\obF_{n\bk,\lambda}
}{
\left[
\omega_{n\bk,\lambda}^2-\omega_{\rm Er}^2
\right]^2
+
\gamma_{n\bk,\lambda}^2\omega_{\rm Er}^2
}
\label{eq:honeycomb_plasmon_loss_suppressed}
\end{equation}
is suppressed by the detuning from the band edges, while the reactive part
\[
\Re\obG_R^{\rm pl}(\omega_{\rm Er})
\simeq
\sum_{n,\bk,\lambda}
\frac{
\obF_{n\bk,\lambda}
}{
\omega_{n\bk,\lambda}^2-\omega_{\rm Er}^2
}
\]
remains finite.  The Er-Er transfer is therefore still resonant between
the donor and acceptor, but the graphene plasmons enter only virtually.
The coherent exchange interaction is
\begin{equation}
J_{DA}
=
-\frac{\mu_0\omega_{\rm Er}^2}{\hbar}
\bd_A^\ast\cdot
\Re\obG_R(\bx_A,\bx_D;\omega_{\rm Er})
\cdot
\bd_D ,
\label{eq:honeycomb_Er_Er_exchange}
\end{equation}
whereas the unwanted plasmonic loss channel is proportional to
\begin{equation}
\Gamma_{\rm pl}
=
\frac{2\mu_0\omega_{\rm Er}^2}{\hbar}
\bd_D^\ast\cdot
\Im\obG_R^{\rm pl}(\bx_D,\bx_D;\omega_{\rm Er})
\cdot
\bd_D .
\label{eq:honeycomb_Er_plasmon_loss}
\end{equation}
Placing \(\omega_{\rm Er}\) inside the complete metagate-induced plasmon
gap suppresses Eq.~\eqref{eq:honeycomb_Er_plasmon_loss}, while retaining
the off-shell reactive Green-function contribution in
Eq.~\eqref{eq:honeycomb_Er_Er_exchange}.

The numerical design in
Figs.~\ref{fig:SI_Er_DLG_honeycomb_bandstructure} and
\ref{fig:SI_Er_DLG_honeycomb_EF_superlattice} uses
\(N_G=20\), an average Fermi energy near \(1~{\rm eV}\), and a strongly
inversion-broken honeycomb landscape with local Fermi energies extending
to values of order \(E_F^{\max}\sim2~{\rm eV}\).  Such high doping is
ambitious for conventional solid-state gates but feasible as an
electrolyte or ion-gel back-gating target.  In monolayer graphene,
\begin{equation}
n(E_F)
=
\frac{1}{\pi}
\left(
\frac{E_F}{\hbar v_F}
\right)^2
\simeq
7.35\times10^{13}
\left(
\frac{E_F}{1~{\rm eV}}
\right)^2
{\rm cm}^{-2},
\label{eq:graphene_density_vs_EF}
\end{equation}
so \(E_F\simeq2~{\rm eV}\) corresponds to
\(n\simeq3\times10^{14}~{\rm cm}^{-2}\), a density scale accessible in
high-density electrolyte-gated graphene~\cite{efetov2010controlling}.  In a
realistic device, an ion-gel or electrolyte back gate can provide the large
global carrier density, while a nearby patterned metagate separated by a
thin dielectric spacer provides the nanoscale honeycomb modulation.

This design should be viewed as a proof-of-principle band-gap engineering
calculation.  A quantitative device model should replace
Eq.~\eqref{eq:honeycomb_DLG_Coulomb_pm} by the full metagate Coulomb
operator, include graphene quantum capacitance, nonlocal response,
dielectric anisotropy, finite metagate thickness, edge disorder, and
Drude damping.  It is also essential to avoid domain walls, edges, or
defect modes inside the active region, because such modes would
intentionally reintroduce in-gap plasmon channels.  For FRET suppression,
the desired structure is a uniform broken-inversion honeycomb plasmonic
crystal with \(\omega_{\rm Er}\) lying in the bulk band gap and no
localized in-gap state coupled to the Er emitters.

\subsection{Rare-earth laser power scaling and directed-energy implications}
\label{subsec:rare_earth_laser_power_scaling_applications}
\label{subsec:ETU_laser_power_scaling}

The same FRET-screening mechanism could have important consequences
beyond spectroscopy and nanoscale energy transfer.  Rare-earth-doped
solids and fibers form a central materials platform for lasers and optical
amplifiers, including Er-, Nd-, Yb-, Tm-, and Ho-doped gain media used in
telecommunications, eye-safer ranging, lidar, medical lasers, industrial
processing, and high-power fiber-laser systems
\cite{Miniscalco1991,Richardson2010HighPowerFiberLasers,
Jauregui2013HighPowerFibreLasers}.  In these media, increasing the
active-ion concentration is highly desirable because it increases pump
absorption, small-signal gain, stored energy density, and the possibility of
shorter or more compact devices.  However, the usable concentration is
often limited by concentration quenching, excitation migration, and
energy-transfer upconversion (ETU), in which resonant energy transfer
between neighboring rare-earth ions converts useful excitation into heat or
population in unwanted manifolds \cite{Auzel2004,Spariosu1994,
Pollnau2003Heat,Kim2008ImpactETU,Kim2009FiberLaserPumpedErYAG}.
For example, in Er-doped laser media two neighboring excited ions can
undergo an ETU process in which one ion relaxes while the other is
promoted to a higher-lying manifold; subsequent multiphonon relaxation
then converts useful stored excitation into heat.  In Er:YAG operating near
the \({}^{4}I_{13/2}\to{}^{4}I_{15/2}\) transition, this process is a major
power-scaling constraint because it shortens the effective upper-state
lifetime, increases the threshold, and adds heat loading
\cite{Kim2008ImpactETU,Kim2009FiberLaserPumpedErYAG}.

The connection to the present theory is direct.  ETU is a pair process whose
microscopic rate is controlled by the same resonant dipole-dipole Green
tensor that governs FRET.  We define the FRET suppression factor of the
structured electromagnetic environment as
\begin{equation}
S_{\rm FRET}
\equiv
\frac{
\Gamma_{\rm FRET}^{\rm env}
}{
\Gamma_{\rm FRET}^{\rm bulk}
}.
\label{eq:S_FRET_definition_laser}
\end{equation}
For the \(N_G=10\) double multilayer-graphene structure shown in
Fig.~\ref{fig:Er_FRET_NG10}, the reflective Dirichlet-like regime can
achieve
\begin{equation}
S_{\rm FRET}\sim 10^{-5},
\label{eq:S_FRET_1e_minus_5}
\end{equation}
corresponding to a suppression of Er-Er FRET by five orders of
magnitude.  In an ideal ETU-limited laser medium, where resonant
dipole-mediated rare-earth-rare-earth transfer is the dominant
concentration-limiting mechanism, the macroscopic ETU coefficient scales
approximately as
\begin{equation}
C_{\rm ETU}^{\rm env}
\simeq
S_{\rm FRET}\,
C_{\rm ETU}^{\rm bulk}.
\label{eq:C_ETU_suppressed}
\end{equation}

To estimate the resulting concentration and power scaling, let
\(N_{\rm RE}\) be the active rare-earth ion density and let
\(N_2=\beta N_{\rm RE}\) be the upper-laser-level population density at a
fixed fractional inversion \(\beta\).  A minimal rate equation contains
\begin{equation}
\frac{dN_2}{dt}
=
R_{\rm p}
-
\frac{N_2}{\tau_2}
-
C_{\rm ETU}N_2^2
+\cdots ,
\label{eq:laser_rate_equation_ETU}
\end{equation}
where \(R_{\rm p}\) is the pump rate density and \(\tau_2\) is the intrinsic
upper-state lifetime in the absence of ETU.  The ETU loss rate per excited
ion is
\begin{equation}
\Gamma_{\rm ETU}^{\rm per\,ion}
=
C_{\rm ETU}N_2
=
C_{\rm ETU}\beta N_{\rm RE}.
\label{eq:ETU_per_ion_rate}
\end{equation}
If the usable concentration is defined by requiring this ETU loss rate to
remain below a fixed fraction \(\xi\) of the intrinsic decay rate,
\begin{equation}
C_{\rm ETU}\beta N_{\rm RE,max}
\lesssim
\frac{\xi}{\tau_2},
\label{eq:ETU_limited_concentration_condition}
\end{equation}
then
\begin{equation}
N_{\rm RE,max}
\sim
\frac{\xi}{\beta C_{\rm ETU}\tau_2}.
\label{eq:Nmax_ETU}
\end{equation}
Combining Eqs.~\eqref{eq:C_ETU_suppressed} and
\eqref{eq:Nmax_ETU} gives the ideal ETU-limited concentration
enhancement
\begin{equation}
\frac{
N_{\rm RE,max}^{\rm env}
}{
N_{\rm RE,max}^{\rm bulk}
}
\simeq
\frac{1}{S_{\rm FRET}}.
\label{eq:ideal_concentration_enhancement}
\end{equation}
If the laser remains ETU-limited rather than pump-brightness,
thermal-extraction, reabsorption, optical-damage, parasitic-nonlinearity,
or resonator-design limited, the stored energy density, small-signal gain,
and extractable power density scale approximately with the usable active-ion
density:
\begin{equation}
\frac{
P_{\rm max}^{\rm env}
}{
P_{\rm max}^{\rm bulk}
}
\sim
\frac{
N_{\rm RE,max}^{\rm env}
}{
N_{\rm RE,max}^{\rm bulk}
}
\simeq
\frac{1}{S_{\rm FRET}}.
\label{eq:ideal_laser_power_enhancement}
\end{equation}
Thus, the five-orders-of-magnitude suppression
\(S_{\rm FRET}\sim10^{-5}\) demonstrated in
Fig.~\ref{fig:Er_FRET_NG10} corresponds to an ideal \(10^5\)-fold
enhancement of the ETU-limited active-ion loading and power-density
bound.  Equivalently, one could aim for the same ETU-limited output power
using an active-medium volume smaller by a factor approaching \(10^5\),
provided the other system-level constraints are also scaled appropriately.

This scaling could be valuable for several classes of rare-earth laser
systems.  In compact eye-safer transmitters based on Er- or Ho-doped
media, suppressing resonant ion-ion transfer would enable higher
rare-earth loading without the usual lifetime penalty, improving pump
absorption and reducing device length.  In integrated photonic amplifiers
and nanophotonic rare-earth lasers, FRET screening could help reconcile
the need for high ion density with the requirement of maintaining a long
excited-state lifetime.  In high-power Yb-doped fiber and slab
architectures, the same concept could complement existing strategies for
raising gain per unit length while controlling parasitic quenching and heat
generation.  In each case, the central advantage is that the active-ion
density would be controlled not only by chemistry and host choice, but
also by the engineered electromagnetic Green tensor of the surrounding
heterostructure.

\paragraph{\textbf{Rangefinding, LiDAR, and laser remote sensing.}}
The same materials-level power-density enhancement would be valuable for
laser rangefinding, LiDAR, and laser remote sensing.  These applications
already rely heavily on rare-earth gain media.  Eye-safer rangefinders
commonly use Er-doped glass transmitters near \(1.54~\mu{\rm m}\), while
Ho- and Tm/Ho-doped crystals near \(2.0\text{-}2.1~\mu{\rm m}\) are
attractive for laser radar, coherent remote sensing, and pumping
mid-infrared optical parametric oscillators
\cite{L3HarrisErGlass,TeledyneFLIRHoYAG,TeledyneFLIRCTHYAG,
Asai1992TmHoYAGLaserRadar,Killinger1992SolidStateLidar}.  In all of
these systems, higher rare-earth loading can increase pulse energy, average
power, or gain per unit length.  Suppressing resonant ion-ion FRET and
the associated concentration quenching would therefore allow either a
brighter transmitter at fixed active-medium volume or a smaller transmitter
at fixed pulse energy.

The effect on maximum range is important but sublinear.  For a hard diffuse
target, the received pulse energy can be written schematically as
\begin{equation}
E_{\rm r}(R)
\propto
E_{\rm tx}
\frac{A_{\rm r}}{R^2}
\frac{1}{R^2}
T_{\rm atm}^2(R),
\label{eq:rangefinder_R4_scaling}
\end{equation}
where \(E_{\rm tx}\) is the transmitted pulse energy, \(A_{\rm r}\) is the
receiver aperture area, and \(T_{\rm atm}(R)\) is the one-way atmospheric
transmission.  The two powers of \(1/R^2\) represent the outgoing beam
spreading and the return collection geometry.  Neglecting atmospheric
extinction and holding receiver sensitivity fixed, the maximum range scales
as
\begin{equation}
\frac{R_{\max}^{\rm env}}{R_{\max}^{\rm bulk}}
\sim
\left(
\frac{E_{\rm tx}^{\rm env}}{E_{\rm tx}^{\rm bulk}}
\right)^{1/4}.
\label{eq:Rmax_fourth_root_scaling}
\end{equation}
Thus an ideal ETU-limited transmitter-energy enhancement of \(10^5\)
would correspond to an ideal hard-target range enhancement of
\(10^{5/4}\simeq17.8\), rather than \(10^5\).  Alternatively, the same
range could be achieved with a substantially smaller active medium, reduced
pump requirement, or improved signal-to-noise margin.

For atmospheric backscatter LiDAR and differential absorption LiDAR
(DIAL), the standard LiDAR equation often contains a \(1/R^2\) geometric
factor for the received backscatter power, together with the
range-dependent backscatter coefficient and two-way atmospheric
transmission.  In the ideal signal-limited case this gives a more favorable
scaling,
\begin{equation}
\frac{R_{\max}^{\rm env}}{R_{\max}^{\rm bulk}}
\sim
\left(
\frac{P_{\rm tx}^{\rm env}}{P_{\rm tx}^{\rm bulk}}
\right)^{1/2},
\label{eq:Rmax_sqrt_lidar_scaling}
\end{equation}
before atmospheric attenuation and background-noise limits are included.
Consequently, a \(10^5\) enhancement of the ETU-limited transmitter
power-density bound could translate, in the ideal signal-limited case, into
a maximum-range enhancement approaching \(10^{5/2}\simeq316\) for
volume-backscatter measurements, before atmospheric extinction,
background, detector, and eye-safety constraints are imposed.  More
conservatively, it could enable the same range and signal-to-noise ratio
with a much smaller and lower-SWaP rare-earth transmitter.

\paragraph{\textbf{Laser power beaming and remote charging.}}
A closely related non-destructive application is laser power beaming, in
which optical power is transmitted from a ground, vehicle, or airborne
platform to a remote photovoltaic receiver.  This approach is being explored
for remote energy delivery, persistent sensing, and in-flight charging of
unmanned aerial vehicles (UAVs) \cite{NASA2003LaserPowerUAV,
RAND2011LaserPowerHAA,DARPA_POWER,DARPA_POWER_2025_Record,
Jaafar2021LaserChargedUAV,Lahmeri2022LaserPowerBeamingUAV}.  The
motivation is straightforward: the endurance of electrically powered UAVs
is often limited by onboard battery mass, whereas a laser power-beaming link
can replenish or supplement the onboard energy store without landing.

The delivered electrical power can be written schematically as
\begin{equation}
P_{\rm elec}^{\rm rec}
=
\eta_{\rm atm}(R)
\eta_{\rm opt}
\eta_{\rm PV}
P_{\rm laser},
\label{eq:power_beaming_received_power}
\end{equation}
where \(P_{\rm laser}\) is the transmitted optical power,
\(\eta_{\rm atm}(R)\) is the range-dependent atmospheric transmission,
\(\eta_{\rm opt}\) accounts for pointing, beam-quality, and optical
collection losses, and \(\eta_{\rm PV}\) is the photovoltaic conversion
efficiency at the receiver.  A rare-earth gain medium whose ETU-limited
power-density bound is enhanced by FRET screening can therefore increase
the available transmitter power, reduce the gain volume needed for a fixed
power-beaming link, or lower the size, weight, and cooling burden of the
laser transmitter.

The range scaling is more favorable than in hard-target LiDAR because
power beaming is a one-way link.  For a diffraction- or divergence-limited
beam with approximately fixed far-field divergence \(\theta\), the beam
area at range \(R\) scales as \(\pi(\theta R)^2\), and the power collected
by a receiver of area \(A_{\rm rec}\) scales approximately as
\begin{equation}
P_{\rm rec}(R)
\propto
P_{\rm laser}
\frac{A_{\rm rec}}{\pi(\theta R)^2}
T_{\rm atm}(R).
\label{eq:power_beaming_R2_scaling}
\end{equation}
Neglecting atmospheric extinction and holding the receiver aperture, beam
divergence, and conversion efficiency fixed, the maximum range for a given
required received power scales as
\begin{equation}
\frac{
R_{\max}^{\rm env}
}{
R_{\max}^{\rm bulk}
}
\sim
\left(
\frac{
P_{\rm laser}^{\rm env}
}{
P_{\rm laser}^{\rm bulk}
}
\right)^{1/2}.
\label{eq:power_beaming_range_scaling}
\end{equation}
Thus an ideal \(10^5\)-fold enhancement of the ETU-limited rare-earth
power-density bound would imply an ideal one-way power-beaming range
enhancement of \((10^5)^{1/2}\simeq316\), before atmospheric turbulence,
pointing jitter, receiver heating, photovoltaic saturation, safety
constraints, and optical-damage limits are included.  At fixed range and
fixed delivered electrical power, the same scaling could instead reduce the
required active gain volume, cooling load, or transmitter mass.

\paragraph{\textbf{Directed-energy implications.}}
The most dramatic extrapolation is to directed-energy laser platforms.
Modern high-energy-laser systems for counter-unmanned-aircraft and
short-range air-defense missions rely heavily on electrically powered
solid-state or fiber-laser architectures and beam combining.  The U.S.
Army's Directed Energy Maneuver-Short Range Air Defense
(DE M-SHORAD) prototype, for example, is described as a Stryker-mounted
system whose primary weapon is a \(50~{\rm kW}\) spectral-beam-combined
laser \cite{DOTE2024DEMSHORAD}.  Navy directed-energy development
materials identify compact, robust fiber lasers as attractive HEL sources,
but also emphasize stimulated Brillouin scattering, thermal mode
instability, and the need for spectral or coherent beam combining as key
power-scaling challenges, with operation across the ytterbium-doped fiber
spectrum \cite{NavySBIR2023YbFiberSBC}.  Other deployed or advertised
counter-UAS HEL systems span the \(10~{\rm kW}\) to \(100~{\rm kW}\) class
\cite{RaytheonHEL,RafaelIronBeam}.  If the gain modules in such
architectures were ETU-limited, then a Green-function-engineered
rare-earth medium with \(S_{\rm FRET}=10^{-5}\) would correspond, at the
materials level, to an ideal \(10^5\)-fold enhancement of the active-ion
concentration and power-density bound.

This should not be interpreted as a direct prediction that a complete
\(50~{\rm kW}\) laser weapon becomes a \(5~{\rm GW}\) deployed system.
Electrical generation, pump diodes, cooling, nonlinear propagation, beam
quality, atmospheric propagation, optical damage, and beam-combining
architecture would impose additional limits.  Nevertheless, even a small
fraction of the ideal materials-level scaling would be consequential: it
could increase output power at fixed active-medium volume, reduce size,
weight, and cooling burden at fixed output power, or move compact
counter-UAS laser modules closer to practical deployment.  The broader
conclusion is that FRET screening offers a materials-level route to
rare-earth laser power scaling, with possible impact ranging from
integrated photonics, rangefinding, LiDAR, remote sensing, and optical
power beaming to defense-relevant high-power laser systems.

Several qualifications are important.  Equation
\eqref{eq:ideal_laser_power_enhancement} is an upper bound: it assumes
that ETU is the dominant concentration-limiting mechanism and that the
structured electromagnetic environment suppresses the relevant pair
transfer rates over the dominant ion-ion separation distribution.  In a
real laser medium, residual concentration quenching may also arise from
nearest-neighbor clustering, direct exchange, defect-assisted relaxation,
multiphonon relaxation, reabsorption, thermal lensing, optical damage,
pump transport, and host-material limitations.  Nevertheless, the scaling
estimate identifies a new design principle: by engineering the retarded
Green tensor of a rare-earth-doped spacer with tunable two-dimensional
conductors, one can in principle decouple active-ion density from the usual
ETU penalty.  This suggests a route toward nanolaminated or
waveguide-integrated rare-earth laser media in which FRET screening
enables active-ion concentrations, stored energies, and laser powers far
beyond conventional ETU-limited values.

\subsection{Atomic clocks and navigation timing}
\label{subsec:atomic_clocks_FRET_screening}

Another important application of FRET screening is precision timekeeping.
Modern atomic clocks achieve their stability by referencing an oscillator to
an extremely narrow atomic, ionic, or nuclear transition
\cite{Ludlow2015OpticalClocks}.  In neutral-atom optical lattice clocks,
many atoms can be interrogated simultaneously, improving the
quantum-projection-noise-limited stability through the usual
\(\sqrt{N}\) enhancement.  However, increasing the number density or
placing emitters closer together can introduce density-dependent
collisional shifts, resonant dipole-dipole shifts, superradiant decay
channels, and dephasing mechanisms that limit clock accuracy and
stability \cite{Chang2004DipoleClockShifts,Swallows2011CollisionalShifts,
Aeppli2024OpticalClock}.  These effects are especially relevant for dense
ensemble clocks, rare-earth-ion solid-state clocks, and proposed
solid-state nuclear clocks, where very large numbers of emitters are
embedded in a crystalline host
\cite{PeikTamm2003ThClock,Higgins2025ThTemperature,
Ooi2026,morgan2025proposal}.  In such systems,
FRET-like resonant dipole-mediated transfer provides a natural
concentration-dependent channel for line broadening, energy migration,
frequency shifts, and loss of coherence.

The present heterostructure provides a possible way to suppress this
clock-density penalty.  Let
\begin{equation}
S_{\rm FRET}
=
\frac{
\Gamma_{\rm FRET}^{\rm env}
}{
\Gamma_{\rm FRET}^{\rm bulk}
}
\label{eq:clock_SFRET_definition}
\end{equation}
be the suppression factor for resonant dipole-mediated transfer in the
structured electromagnetic environment.  For the multilayer-graphene
configuration in Fig.~\ref{fig:Er_FRET_NG10}, the calculated suppression
can reach
\begin{equation}
S_{\rm FRET}\sim 10^{-5}.
\label{eq:clock_SFRET_1e5}
\end{equation}
If the relevant clock transition is limited by such resonant
dipole-mediated dephasing or frequency shifts, the corresponding
FRET-limited linewidth or shift scales as
\begin{equation}
\Delta\nu_{\rm dd}^{\rm env}
\simeq
S_{\rm FRET}
\Delta\nu_{\rm dd}^{\rm bulk},
\label{eq:clock_dd_shift_suppression}
\end{equation}
where \(\Delta\nu_{\rm dd}\) denotes the dipole-mediated contribution to
the clock linewidth, shift, or dephasing rate.  Therefore, in the ideal
FRET-limited regime,
\begin{equation}
\frac{
\Delta\nu_{\rm dd}^{\rm bulk}
}{
\Delta\nu_{\rm dd}^{\rm env}
}
\simeq
\frac{1}{S_{\rm FRET}}
\sim
10^5 .
\label{eq:clock_ideal_linewidth_enhancement}
\end{equation}
This is the direct analogue of the ETU-limited rare-earth laser scaling:
the same Green-function engineering that suppresses resonant energy
transfer also suppresses the corresponding clock-decoherence channel.

There is a second, independent advantage.  If resonant dipole-dipole
interactions limit the usable emitter density, then the maximum number of
clock emitters can ideally be increased by the same factor,
\begin{equation}
\frac{
N_{\max}^{\rm env}
}{
N_{\max}^{\rm bulk}
}
\simeq
\frac{1}{S_{\rm FRET}}
\sim
10^5 .
\label{eq:clock_Nmax_enhancement}
\end{equation}
For a clock operating at the standard quantum limit, the fractional
frequency instability scales approximately as
\begin{equation}
\sigma_y(\tau)
\sim
\frac{1}{2\pi\nu_0 T}
\sqrt{
\frac{T_c}{N\tau}
},
\label{eq:clock_SQL_scaling}
\end{equation}
where \(\nu_0\) is the clock frequency, \(T\) is the interrogation time,
\(T_c\) is the experimental cycle time, \(N\) is the number of interrogated
emitters, and \(\tau\) is the averaging time.  If FRET screening only
allows a larger usable emitter number while all other clock parameters
remain fixed, then
\begin{equation}
\frac{
\sigma_y^{\rm bulk}
}{
\sigma_y^{\rm env}
}
\simeq
\sqrt{
\frac{N_{\max}^{\rm env}}{N_{\max}^{\rm bulk}}
}
\simeq
\frac{1}{\sqrt{S_{\rm FRET}}}
\sim
316 .
\label{eq:clock_SQL_316}
\end{equation}
If, in addition, the interrogation time itself is limited by the same
dipole-mediated dephasing channel, then suppressing that dephasing by
\(10^5\) can also increase the usable coherence time.  In the ideal case
where both \(N\) and \(T\) are FRET-limited, the clock stability
enhancement can approach
\begin{equation}
\frac{
\sigma_y^{\rm bulk}
}{
\sigma_y^{\rm env}
}
\sim
\sqrt{
\frac{N_{\max}^{\rm env}}{N_{\max}^{\rm bulk}}
\frac{T^{\rm env}}{T^{\rm bulk}}
}
\sim
10^5 .
\label{eq:clock_ideal_1e5_stability}
\end{equation}
Thus, depending on whether FRET limits the emitter number alone, the
coherence time alone, or both simultaneously, the ideal clock improvement
ranges from a factor of order \(316\) to \(10^5\).

This scaling is particularly suggestive for compact solid-state clocks.
Rare-earth-ion-doped crystals and thorium-doped crystals can host many
more emitters than dilute free-space atomic ensembles, but the same high
density that improves signal strength can introduce inhomogeneous
broadening, ion-ion interactions, spectral diffusion, and
concentration-dependent shifts \cite{Higgins2025ThTemperature,
Ooi2026,morgan2025proposal,
Pignol2024DipoleDecoherence}.  A nanolaminated clock medium placed
between tunable two-dimensional conductors would add a new
electromagnetic design knob: rather than reducing the emitter density to
avoid dipole-mediated broadening, one could suppress the nonlocal Green
tensor responsible for the unwanted transfer channel.  This could enable a
dense, compact, and robust clock material with both high signal-to-noise
ratio and reduced density-dependent dephasing.

The implications for satellite navigation and GPS-like timing systems are
also clear, although they must be interpreted carefully.  GPS uses atomic
clocks on satellites to provide precise timing signals to receivers
\cite{Ashby2003GPSRelativity,GPSGovTime}.  The GPS Standard Positioning
Service performance standard specifies a time-transfer error of
\(\le 30~{\rm ns}\), 95\% of the time, for the signal-in-space timing
component under representative user conditions \cite{GPSSPS2020}.  Since
a timing error \(\Delta t\) corresponds to a range-equivalent error
\(c\Delta t\), a \(30~{\rm ns}\) timing error corresponds to approximately
\(9~{\rm m}\).  If this timing uncertainty were purely clock-limited and
if the clock error channel were improved by \(10^5\), the timing error
would become
\begin{equation}
30~{\rm ns}
\longrightarrow
0.3~{\rm ps},
\label{eq:GPS_30ns_to_03ps}
\end{equation}
corresponding to a range-equivalent timing error
\begin{equation}
c(0.3~{\rm ps})
\simeq
0.09~{\rm mm}.
\label{eq:GPS_009mm}
\end{equation}
This is an ideal clock-limited upper bound, not a prediction that deployed
GPS positioning would automatically improve by \(10^5\).  In practice,
satellite navigation accuracy is also limited by ephemeris and orbit
errors, ionospheric and tropospheric propagation delays, multipath,
receiver noise, antenna phase-center effects, clock synchronization across
the constellation, and geometry dilution of precision
\cite{Ashby2003GPSRelativity,GPSSPS2020}.  Nevertheless, reducing the
clock-limited timing channel from nanoseconds to the sub-picosecond regime
would be transformative for precision navigation, synchronization,
distributed sensing, geodesy, and GPS-denied timing architectures.  The
main conclusion is that FRET screening could provide a materials-level
route to denser and more stable clock ensembles, with ideal improvements
ranging from \(\sim316\) in atom-number-limited stability to \(10^5\) in a
fully FRET-limited clock.

\section{Conclusions}\label{sec:conclusion}

We have translated the QED interaction-tuning framework of the conductor-dielectric-conductor heterostructure to spontaneous emission and FRET.  The central mathematical objects are the vacuum propagator and the retarded Maxwell Green tensor.  The Feynman propagator is the natural perturbative object, but spontaneous emission isolates its vacuum/Hadamard component, while FRET reorganizes the second-order time-ordered amplitude into a retarded Green tensor and an absorptive acceptor response.  In the Er-graphene realization, this framework connects the experimentally demonstrated strong dissipative Er-to-graphene interaction \cite{Cano2020ErGraphene} to a complementary dispersive regime in which virtual graphene plasmons mediate Er-Er coupling while on-shell graphene loss is suppressed.

The two-sheet geometry makes these propagators programmable.  The reduced Green function is a one-dimensional multiple-reflection problem in $z$, with a universal denominator $1-r_{\lambda}e^{\I\beta W}$ at real frequency and $1-r_{\lambda}e^{-QW}$ at imaginary or evanescent frequency.  In the near-field TM sector, this denominator generates the same image lattice and Poisson-resummed transverse harmonics found in the static QED interaction problem.  Consequently, FRET can be switched from the bulk $x_\rho^{-6}$ law to an exponentially screened law in the $r_{\TM}\to-1$ branch, or enhanced over a quasi-two-dimensional window in the $r_{\TM}\to1^{-}$ branch.  Because $r_{\TM/\TE}(q_\rho,\omega)$ is gate-tunable in two-dimensional conductors and can be engineered in high-impedance metasurfaces, the platform offers a material-agnostic route to programmable energy transfer in quantum materials, molecular photonics, and excitonic devices.

\begin{acknowledgements}
This work has been supported by the Office of Naval Research (ONR) through the U.S. Naval Research Laboratory (NRL). M. N. L. acknowledges support by the Air Force Office of Scientific Research (AFOSR) under award no. FA9550-23-1-0472.
\end{acknowledgements}

\appendix

\begin{figure}[htb]
\centering
\includegraphics[width=\linewidth]{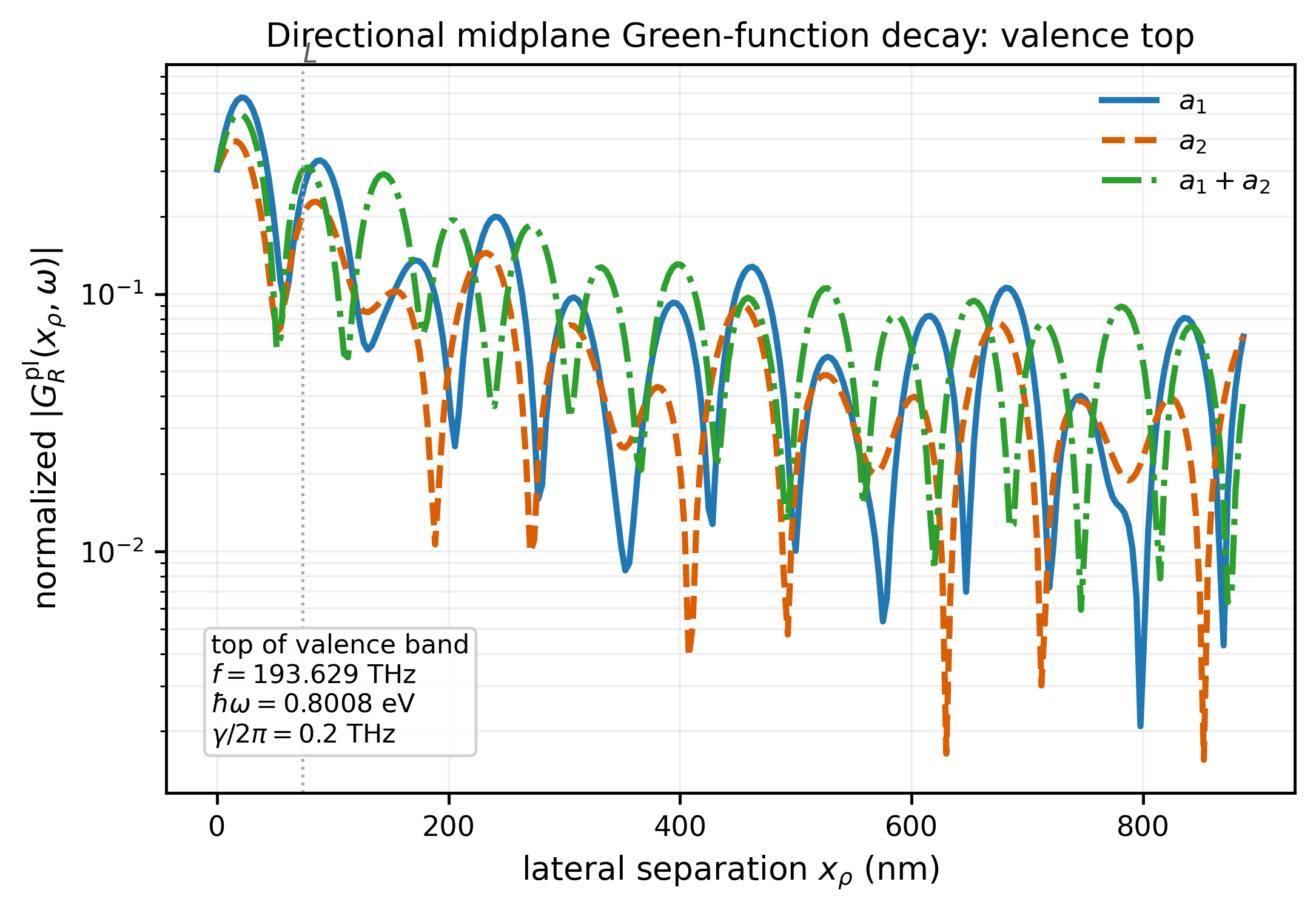}
\caption{
\textbf{Directional Green-function envelope at the top of the lower plasmon band.}
Normalized envelope of the plasmon-mediated midplane retarded Green
function along \(\textbf{a}_1\), \(\textbf{a}_2\),
and \(\textbf{a}_1+\textbf{a}_2\), evaluated at the top of
the lower plasmon band.  Because the probe frequency coincides with a band
edge, the response contains nearly propagating Bloch-plasmon contributions
and is more extended than the in-gap response shown in
Fig.~\ref{fig:app_green_envelope_Er_transition}.
}
\label{fig:app_green_envelope_valence_top}
\end{figure}

\section{Transition operator, Feynman propagator, and the retarded FRET amplitude}\label{app:Tmatrix}

\begin{figure*}[htb]
\centering
\includegraphics[width=0.96\textwidth]{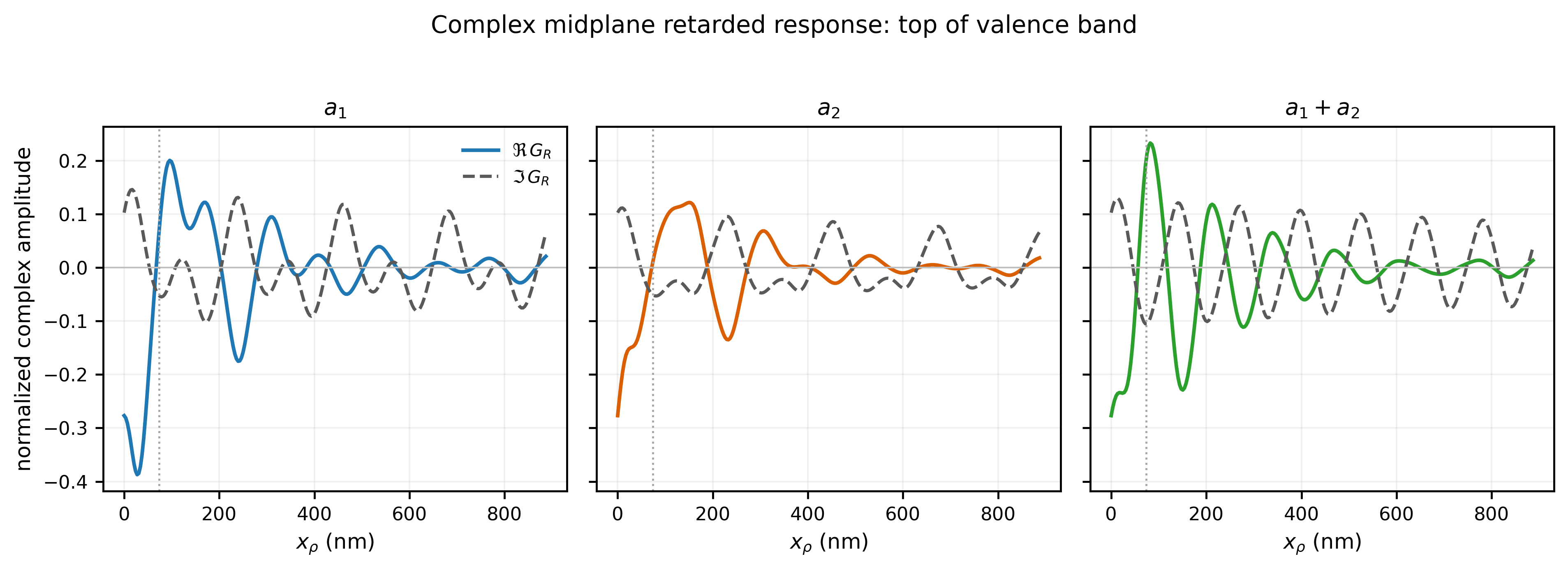}
\caption{
\textbf{Complex directional Green function at the top of the lower plasmon band.}
Real and imaginary parts of the illustrative equal-weight complex response
\(G_{\rm eq}=G_{xx}^{\rm sym}+G_{zz}^{\rm asym}\) along \(\textbf{a}_1\), \(\textbf{a}_2\),
and \(\textbf{a}_1+\textbf{a}_2\), evaluated at the top of the lower plasmon band.  The
slower spatial decay and finite phase oscillations reflect the proximity
to propagating Bloch-plasmon states at the band edge.
}
\label{fig:app_green_complex_valence_top}
\end{figure*}

The interaction-picture time-evolution operator obeys
\begin{equation}
 \I\hbar\frac{\dd}{\dd t}\hat U(t,t_0)=\hat V(t)\hat U(t,t_0),
\end{equation}
with formal solution
\begin{equation}
 \hat U(t,t_0)=1-\frac{\I}{\hbar}\int_{t_0}^{t}\dd t_1\hat V(t_1)\hat U(t_1,t_0).
\end{equation}
Iterating gives the transition operator
\begin{equation}
 \hat T(E)=\hat V+\hat V\frac{1}{E-\hat H_0+\I0^+}\hat V+\cdots.
\end{equation}
For two subsystems $V=V_D+V_A$, the second-order cross terms are
\be
 \hat T_{AD}^{(2)}=V_A\hat G_0(E)V_D+V_D\hat G_0(E)V_A,
\ee
with
\be
 \hat G_0(E)=\frac{1}{E-\hat H_0+\I0^+}.
\ee
Using $V_a=-\hat{\bd}_a\cdot\hat{\bE}(\bx_a)$, the cross term contains field matrix elements of the form
\begin{equation}
 \langle0|\hat E_i(\bx_A)\hat G_0(E)\hat E_j(\bx_D)|0\rangle
 +\langle0|\hat E_j(\bx_D)\hat G_0(E)\hat E_i(\bx_A)|0\rangle.
\end{equation}
This is the energy-domain version of the time-ordered electric Feynman propagator.  Explicitly,
\begin{equation}
 \mathcal D^{EE}_{F,ij}(x,x')=-\I\langle0|T\hat E_i(x)\hat E_j(x')|0\rangle.
\end{equation}
After summing over intermediate photon modes, the two time orderings combine into Eq.~\eqref{eq:FRET_amplitude_GR}, the retarded Green-tensor coupling.  This is why the final FRET rate contains $\obG_{\text R}\Im\,\alpha_A\obG_{\text A}$ rather than a bare standing-wave Feynman propagator.

\begin{figure}[htb]
\centering
\includegraphics[width=\linewidth]{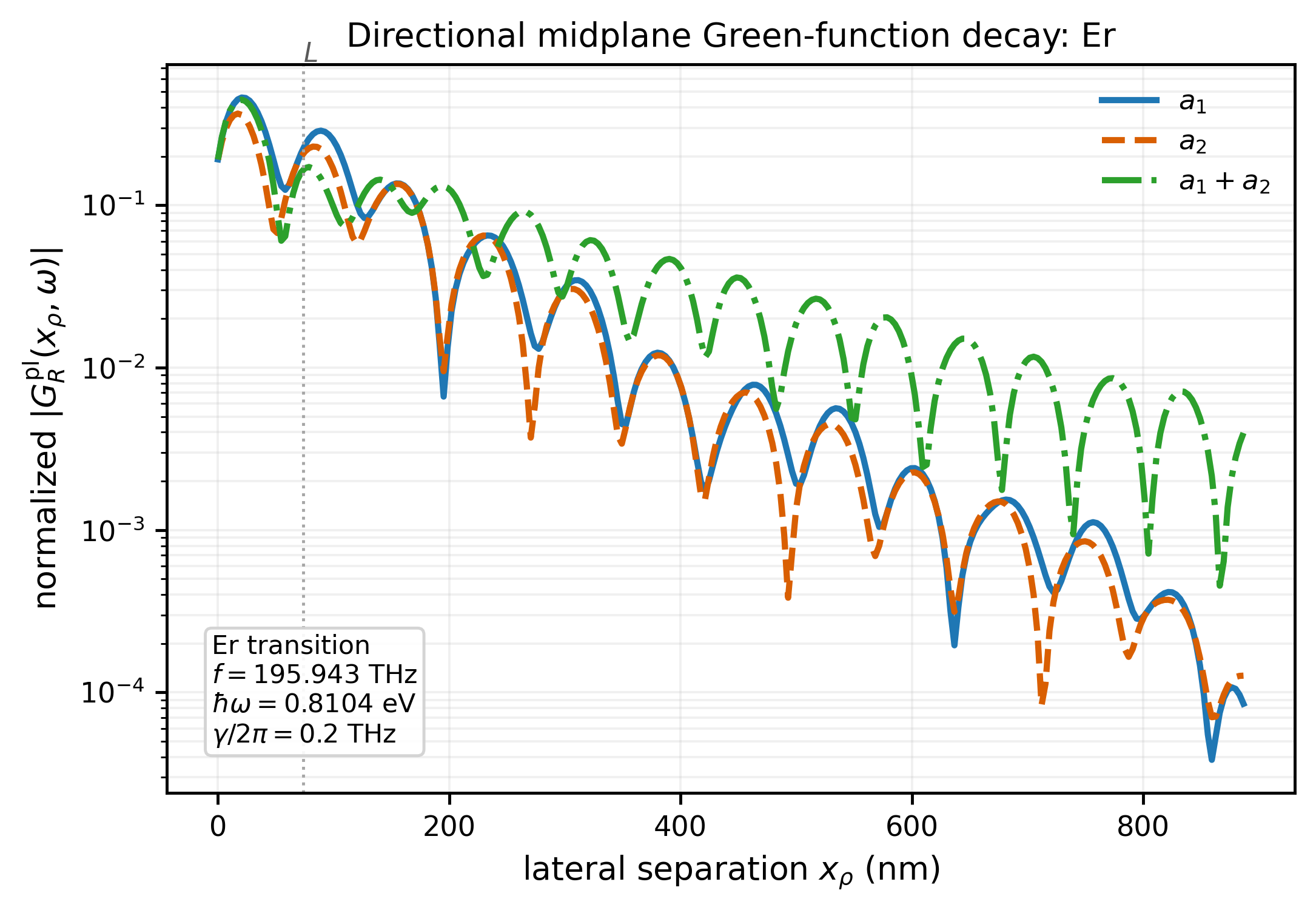}
\caption{
\textbf{Directional Green-function envelope at the Er transition inside the plasmonic gap.}
Normalized envelope of the plasmon-mediated midplane retarded Green
function along \(\textbf{a}_1\), \(\textbf{a}_2\),
and \(\textbf{a}_1+\textbf{a}_2\), evaluated at
\(\omega_{\rm Er}\).  Since the Er transition lies inside the complete
metagate-induced plasmonic band gap shown in
Fig.~\ref{fig:SI_Er_DLG_honeycomb_bandstructure}, there is no on-shell
propagating graphene-plasmon Bloch mode.  The Green function is therefore
evanescent and decays approximately exponentially with the lateral
separation \(x_\rho\).
}
\label{fig:app_green_envelope_Er_transition}
\end{figure}

\begin{figure*}[htb]
\centering
\includegraphics[width=0.96\textwidth]{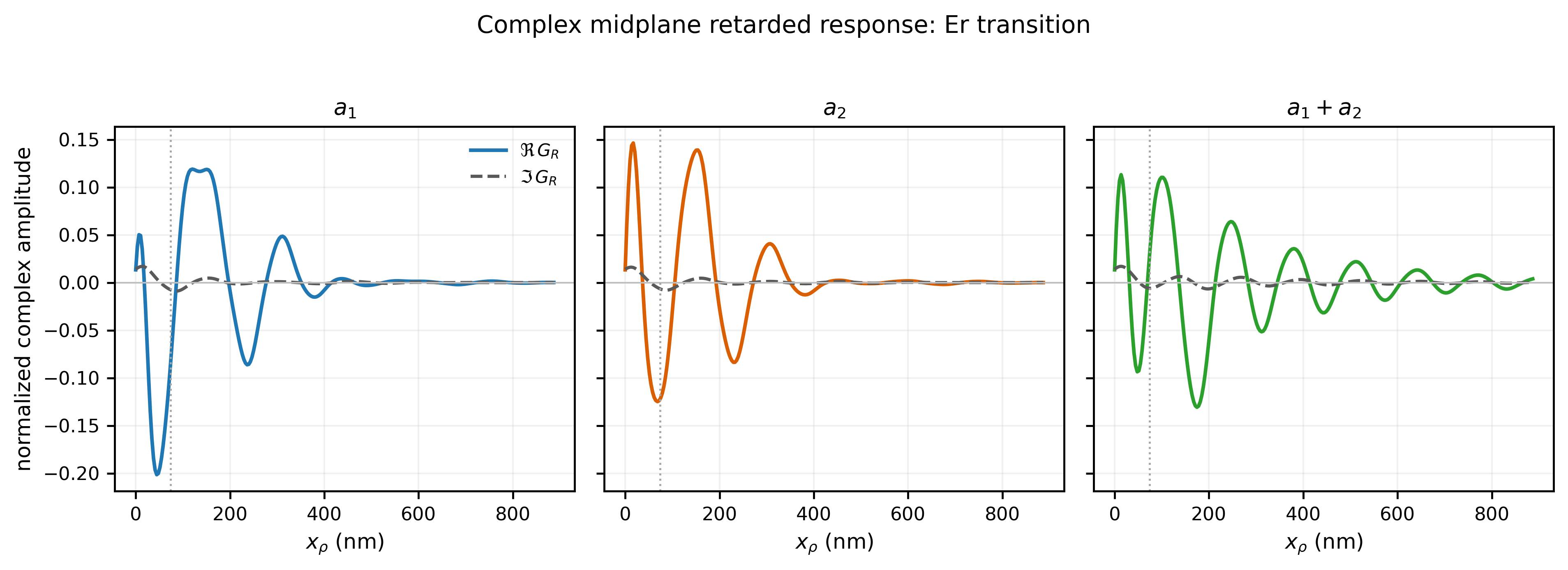}
\caption{
\textbf{Complex directional Green function at the Er transition.}
Real and imaginary parts of the illustrative equal-weight complex response
\(G_{\rm eq}=G_{xx}^{\rm sym}+G_{zz}^{\rm asym}\) along \(\textbf{a}_1\), \(\textbf{a}_2\),
and \(\textbf{a}_1+\textbf{a}_2\), evaluated at \(\omega_{\rm Er}\).  The response is
off-resonant with all propagating graphene-plasmon Bloch modes and is
therefore dominated by evanescent virtual-plasmon contributions.
}
\label{fig:app_green_complex_Er_transition}
\end{figure*}

\section{Contour prescriptions and cylindrical waves}
\label{app:contours}

At fixed positive frequency in two spatial dimensions, the retarded and
advanced scalar Helmholtz Green functions are determined by their pole
prescriptions.  With the scalar normalization used in this work,
\begin{align}
G_R^{2D}(x_\rho;\omega)
&=
-\int\frac{\dd^2\bq_\rho}{(2\pi)^2}
\frac{e^{\I\bq_\rho\cdot\brho}}
{k^2-q_\rho^2+\I0^+}
=
\frac{\I}{4}H_0^{(1)}(kx_\rho),
\label{eq:GR_2D_Hankel}
\\
G_A^{2D}(x_\rho;\omega)
&=
-\int\frac{\dd^2\bq_\rho}{(2\pi)^2}
\frac{e^{\I\bq_\rho\cdot\brho}}
{k^2-q_\rho^2-\I0^+}
=
-\frac{\I}{4}H_0^{(2)}(kx_\rho).
\label{eq:GA_2D_Hankel}
\end{align}
The retarded function is outgoing and the advanced function is incoming.
Their spectral difference and time-symmetric average are
\begin{align}
G_R^{2D}-G_A^{2D}
&=
\frac{\I}{2}J_0(kx_\rho)
=
2\I\,\Im G_R^{2D},
\label{eq:GR_minus_GA_2D}
\\
\frac12\left(G_R^{2D}+G_A^{2D}\right)
&=
-\frac14Y_0(kx_\rho).
\label{eq:GR_plus_GA_2D}
\end{align}
Thus \(\I J_0/2\) is the retarded-advanced spectral difference, not the
Feynman propagator.  At zero temperature the lesser propagator has support
only at negative frequency.  Therefore, for \(\omega>0\),
\begin{equation}
D_F(\omega)
=
D_R(\omega)+D^<(\omega)
=
D_R(\omega).
\label{eq:DF_equals_DR_positive_frequency}
\end{equation}
FRET perturbation theory begins with the time-ordered exchange propagator, but
after the two virtual-photon time orderings are combined, the physical
donor-to-acceptor propagation is governed by the outgoing retarded Green
function.

\section{Poisson summation for the static image lattice}\label{app:poisson}

Starting from
\begin{equation}
 S(\eta;r)=\sum_{m=-\infty}^{\infty}\frac{r^{|m|}}{\sqrt{\eta^2+m^2}},
\end{equation}
write $r=e^{-a+\I\phi}$ with the branch $\phi=0$ for $r>0$ and $\phi=\pi$ for $r<0$.  For the perfect-reflector limits $a=0$, Poisson summation gives
\begin{equation}
 S(\eta;\phi)=\sum_{n=-\infty}^{\infty}\int_{-\infty}^{\infty}\dd x\,
 \frac{e^{\I\phi |x|}e^{-2\pi\I n x}}{\sqrt{\eta^2+x^2}}.
\end{equation}
Using the evenness of the integrand and the identity
\begin{equation}
 \int_0^{\infty}\frac{\cos(kx)}{\sqrt{x^2+\eta^2}}\dd x=K_0(|k|\eta),
\end{equation}
one obtains the branch-specific results
\begin{align}
 S(\eta;-1)&=2\sum_{\ell=0}^{\infty}K_0[(2\ell+1)\pi\eta],\\
 S(\eta;+1)&=\frac{1}{\eta}+2\sum_{\ell=1}^{\infty}K_0(2\pi\ell\eta).
\end{align}
For $r=e^{-a}$ with $a\ll1$, the zero-mode part is
\begin{equation}
 S_0(\eta;r)\simeq 2K_0(a\eta)
 \simeq 2\left[\ln\left(\frac{2}{a\eta}\right)-\gamma_E\right],
\end{equation}
valid for $a\eta\ll1$.  This gives Eq.~\eqref{eq:log_antiscreening}.

A numerically stable exact form follows from the Schwinger representation
\begin{equation}
 \frac{1}{\sqrt{\eta^2+x^2}}=\frac{1}{\sqrt{\pi}}\int_0^{\infty}\dd s\,s^{-1/2}e^{-s(\eta^2+x^2)}.
\end{equation}
The half-line Gaussian integral
\begin{equation}
 \int_0^{\infty}\dd x\,e^{-sx^2+b x}
 =\frac{\sqrt{\pi}}{2\sqrt{s}}\exp\left(\frac{b^2}{4s}\right)
 \operatorname{erfc}\left(-\frac{b}{2\sqrt{s}}\right)
\end{equation}
with $b=\ln r+\I 2\pi\ell$ gives
\begin{equation}
 I_{\ell}(\eta;r)=\frac{1}{4}\int_0^{\infty}\frac{\dd s}{s}e^{-s\eta^2}
 \sum_{\sigma=\pm}
 \exp\left(\frac{b_{\ell\sigma}^2}{4s}\right)
 \operatorname{erfc}\left(-\frac{b_{\ell\sigma}}{2\sqrt{s}}\right),
\end{equation}
where $b_{\ell\sigma}=\ln r+\I\sigma 2\pi\ell$.  For computation one uses $\operatorname{erfcx}(z)=e^{z^2}\operatorname{erfc}(z)$ to avoid overflow.

\section{Dyadic derivatives and orientation factors}\label{app:dyadic}

The static dipole propagator follows from
\begin{equation}
 D^{\mathrm{stat}}_{ij}(x_\rho)=\partial_i\partial_jD_{\mathrm{stat}}(x_\rho).
\end{equation}
Applying the derivatives to the image lattice gives
\begin{align}
 D^{\mathrm{stat}}_{ij}(x_\rho)&=\frac{1}{4\pi\varepsilon_0\varepsilon_c}
 \sum_{m=-\infty}^{\infty}
 \frac{r^{|m|}}{(x_\rho^2+m^2W^2)^{3/2}}
 \nn\\
&\times \left(3\hat s_{m,i}\hat s_{m,j}-\delta_{ij}\right),
\end{align}
with
\begin{equation}
 \hat{\mathbf s}_m=\frac{\brho+mW\hbz}{\sqrt{x_\rho^2+m^2W^2}}.
\end{equation}
For two dipoles,
\begin{equation}
 V_{dd}=d_{A,i}D_{ij}^{\mathrm{stat}}d_{D,j}.
\end{equation}
In the transparent limit this reduces to the usual
\begin{equation}
 V_{dd}^{(0)}=\frac{1}{4\pi\varepsilon_0\varepsilon_c x_\rho^3}
 \left[3(\bd_A\cdot\hat{\brho})(\bd_D\cdot\hat{\brho})-
 \bd_A\cdot\bd_D\right].
\end{equation}
For out-of-plane interlayer-exciton dipoles, $\bd_a=d_{a,z}\hbz$, the transparent term is negative and scales as $-d_{A,z}d_{D,z}/(4\pi\varepsilon_0\varepsilon_c x_\rho^3)$, while the reflective image terms control the universal crossover.

\section{Finite temperature, loss, and MQED normalization}\label{app:finiteT}

In absorbing media the Green tensor is complex and the electromagnetic field must be quantized with noise-current operators.  The MQED field correlation is
\begin{align}
 \left\langle \hat E_i(\bx,\omega)\hat E_j^{\dagger}(\bx',\omega')\right\rangle_T
 &=\frac{\hbar\mu_0}{\pi}\omega^2
 \left[n_T(\omega)+1\right]
 \nn\\
& \times
\Im G_{\text R,ij}(\bx,\bx';\omega)
 \delta(\omega-\omega'),
\end{align}
with $n_T(\omega)=[\exp(\hbar\omega/k_BT)-1]^{-1}$.  The finite-temperature symmetric propagator is
\begin{equation}
 \mathcal D^{EE}_{\mathrm{sym}}(\omega)=\hbar\mu_0\omega^2
 \coth\left(\frac{\hbar\omega}{2k_BT}\right)
 \Im\,\obG_{\text R}(\omega).
\end{equation}
Spontaneous emission uses the positive-frequency $n_T+1$ part; absorption uses the $n_T$ part.  FRET at finite temperature can include thermally stimulated transfer and back-transfer by replacing the zero-temperature acceptor absorption with the appropriate thermal response function.  The geometry-dependent propagation remains contained in $\obG_{\text R}$ and therefore in the same reflection amplitudes $r_{\TM/\TE}(q_\rho,\omega)$.

\section{Design rules for figures and numerical evaluation}\label{app:numerics}

A compact numerical workflow is:
\begin{enumerate}
\item Choose $W$, $\varepsilon_c(\omega)$, donor frequency $\omega_D$, donor dipole $\bd_D$, and acceptor absorption tensor $\Im\,\balpha_A(\omega_D)$.
\item Choose or calculate $r_{\TM/\TE}(q_\rho,\omega_D)$.  For near-field transfer, $q_\rho\sim 1/x_\rho$ is often much larger than $k_c$, so the evanescent TM response is dominant.
\item Evaluate the reduced propagator Eq.~\eqref{eq:midplane_reduced_retarded} and the dyadic Weyl integral.
\item Insert $\obG_{\text R}$ into Eq.~\eqref{eq:FRET_rate_general}.
\item For static or quasi-static checks, compare against the image-lattice formula Eq.~\eqref{eq:static_image_lattice} and the Bessel limits Eqs.~\eqref{eq:S_minus_Bessel}-\eqref{eq:log_antiscreening}.
\end{enumerate}
For plotting normalized universal curves, one may use the dimensionless ratios
\begin{equation}
 \mathcal R_V(\eta;r)=\left|\frac{\mathcal V_{DA}(\eta;r)}{\mathcal V_{DA}(\eta;0)}\right|,
 \qquad
 \mathcal R_{\Gamma}(\eta;r)=\left|\frac{\mathcal V_{DA}(\eta;r)}{\mathcal V_{DA}(\eta;0)}\right|^2.
\end{equation}
The second ratio is the FRET ON/OFF contrast when donor and acceptor spectra are unchanged by the gate.

\begin{figure}[htb]
\centering
\includegraphics[width=\linewidth]{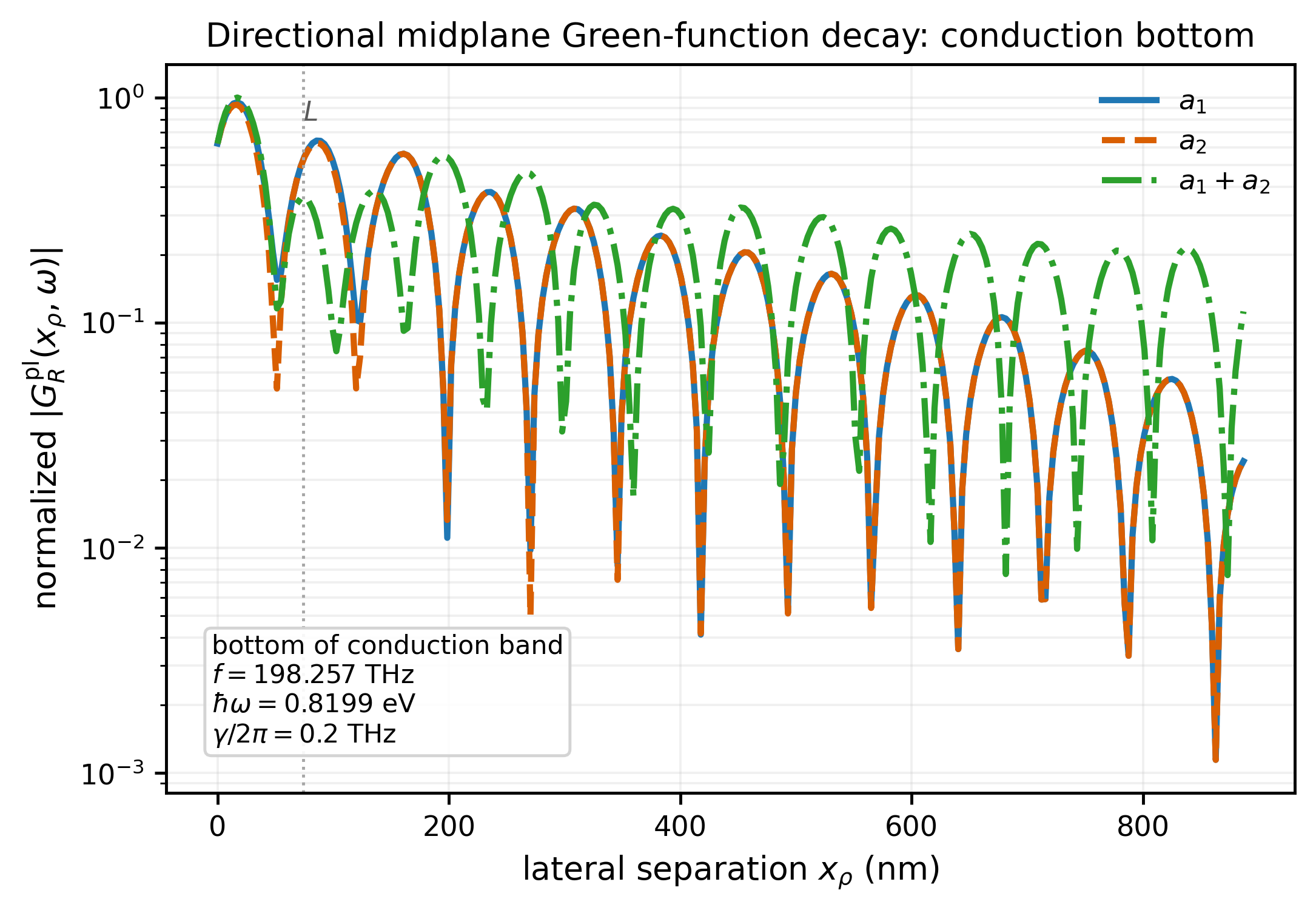}
\caption{
\textbf{Directional Green-function envelope at the bottom of the upper plasmon band.}
Normalized envelope of the plasmon-mediated midplane retarded Green
function along \(\textbf{a}_1\), \(\textbf{a}_2\),
and \(\textbf{a}_1+\textbf{a}_2\), evaluated at the bottom
of the upper plasmon band.  As at the top of the lower band, the probe
frequency approaches a real plasmonic Bloch state, so the response is more
extended than the Er-in-gap Green function.
}
\label{fig:app_green_envelope_conduction_bottom}
\end{figure}

\begin{figure*}[htb]
\centering
\includegraphics[width=0.96\textwidth]{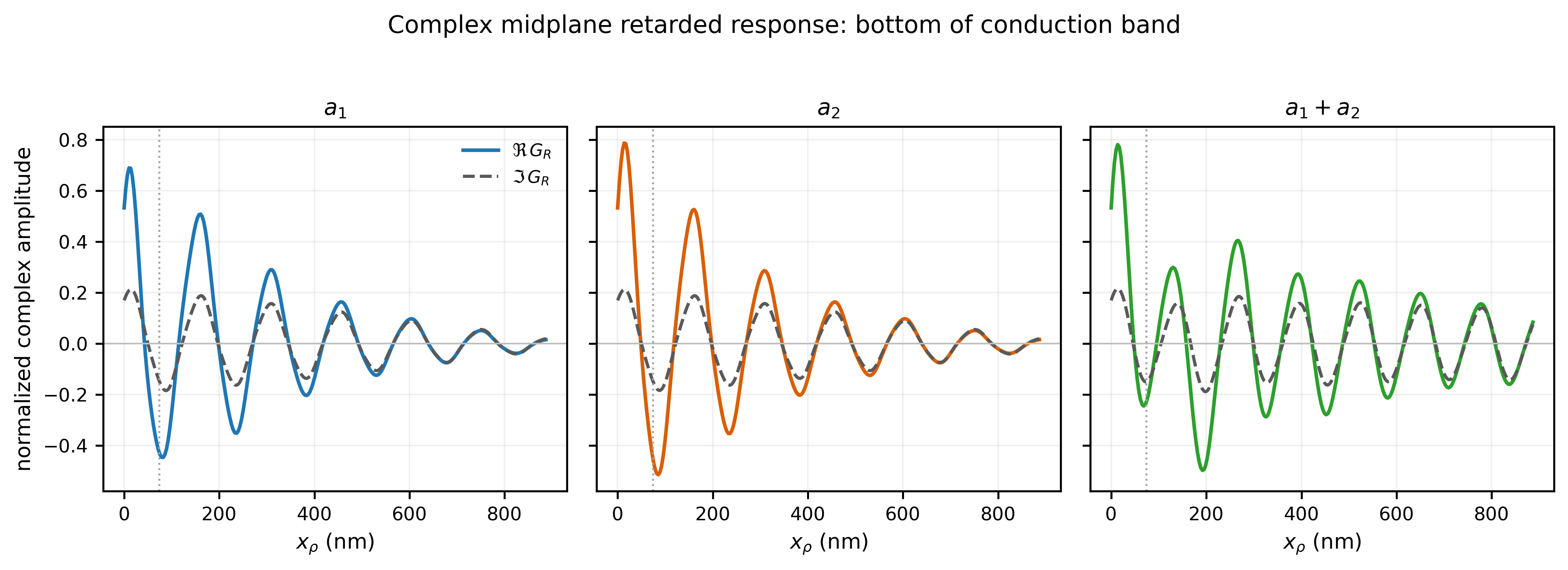}
\caption{
\textbf{Complex directional Green function at the bottom of the upper plasmon band.}
Real and imaginary parts of the illustrative equal-weight complex response
\(G_{\rm eq}=G_{xx}^{\rm sym}+G_{zz}^{\rm asym}\) along \(\textbf{a}_1\), \(\textbf{a}_2\),
and \(\textbf{a}_1+\textbf{a}_2\), evaluated at the bottom of the upper plasmon band.  The
longer-ranged response compared with
Fig.~\ref{fig:app_green_complex_Er_transition} illustrates the contrast
between band-edge propagation and in-gap evanescence.
}
\label{fig:app_green_complex_conduction_bottom}
\end{figure*}

\section{Green functions for metagate-tuned double-layer graphene}
\label{app:metagate_tuned_DLG_green_functions}

In this Appendix we describe how the real-space retarded Green function is
computed from the metagate-tuned double-layer graphene plasmon modes.  The
device concept, the honeycomb broken-inversion Fermi-energy landscape, the
effective multilayer Drude-weight model, and the resulting plasmonic band
gap are introduced in Sec.~\ref{subsec:Er_FRET_metagate_gap}.  The
corresponding real-space Fermi-energy pattern and plasmonic band structure
are shown in
Figs.~\ref{fig:SI_Er_DLG_honeycomb_EF_superlattice} and
\ref{fig:SI_Er_DLG_honeycomb_bandstructure}.  We therefore do not repeat
the metagate construction here.  Instead, we use the symmetric and antisymmetric
Bloch plasmon modes obtained from Eq.~\eqref{eq:honeycomb_DLG_eigenproblem}
to evaluate the spatially resolved midplane Green function.

The purpose of the calculation is to distinguish the in-gap and band-edge
propagation regimes.  When the probe frequency lies inside the complete
plasmonic band gap, there is no real Bloch wave vector satisfying
\(\omega_{n\bk,\lambda}=\omega\), and the plasmon-mediated Green function
is evanescent.  By contrast, when the probe frequency is moved to the top
of the lower plasmon band or to the bottom of the upper plasmon band,
propagating or nearly propagating Bloch plasmons are available and the
Green function becomes much more extended.

\subsection{Mode-resolved retarded Green function}
\label{app:mode_resolved_green_function}

We compute the plasmonic-crystal Bloch modes by extending the plane-wave
formulation for graphene sheets described by Gon\c{c}alves and
Peres~\cite{GoncalvesPeres2016} to the case of plasmonic crystals, i.e. by adapting to the
metagate-defined double-layer graphene (DLG) geometry.  In this approach,
the spatially periodic Fermi-energy landscape \(E_F(\brho)\), and hence
the local Drude weight, is expanded in reciprocal-lattice vectors of the
metagate superlattice.  The electrostatic potential of a plasmonic Bloch
mode is expanded as
\[
\Phi_{n\bk,\lambda}(\brho)
=
\sum_\alpha
\phi^{(n\bk,\lambda)}_\alpha
e^{\I(\bk+\mathbf G_\alpha)\cdot\brho},
\]
where \(\bk\) lies in the first Brillouin zone of the plasmonic crystal.
Solving the resulting plane-wave eigenvalue problem gives the plasmon
frequency \(\omega_{n\bk,\lambda}\) and the Bloch-function coefficients
\(\phi^{(n\bk,\lambda)}_\alpha\).  The method is the direct analogue of
the Bloch-mode construction in the textbook treatment of periodically
modulated graphene plasmons, with the single-layer Coulomb propagator replaced
by the DLG Coulomb propagators appropriate for the symmetric and antisymmetric
layer-parity sectors.

For each Bloch wave vector \(\bk\) and sector \(\lambda=\pm\), the
plasmonic-crystal eigenproblem therefore gives a frequency
\(\omega_{n\bk,\lambda}\) and a set of plane-wave potential coefficients
\(\phi^{(n\bk,\lambda)}_\alpha\).  The \(+\) sector denotes the symmetric double-layer plasmon, while the \(-\) sector denotes the
antisymmetric double-layer plasmon.  We write
\be
\bQ_\alpha=\bk+\mathbf G_\alpha,
\qquad
q_\alpha=|\bQ_\alpha|.
\ee
The plasmon-mediated part of the retarded Green function is evaluated as a
modal spectral sum over these Bloch modes,
\begin{equation}
G_{\eta\eta'}^{\rm pl}(\brho,\brho';\omega)
=
\sum_{\lambda=\pm}
\sum_n
\int_{\rm BZ}\frac{d^2k}{(2\pi)^2}
\frac{
\mathcal F_{n\bk,\lambda}^{(\eta)}(\brho)
\mathcal F_{n\bk,\lambda}^{(\eta')\ast}(\brho')
}{
\omega_{n\bk,\lambda}^2
-
\omega^2
-
\I\gamma_{\rm pl}\omega
}.
\label{eq:app_green_mode_sum_reduced}
\end{equation}
Here \(\eta,\eta'\) label field components and
\(\gamma_{\rm pl}\) is a phenomenological plasmon damping rate.  The
figures below use
\(
\gamma_{\rm pl}/2\pi=0.25~{\rm THz},
\)
which we treat as a low-temperature, low-disorder benchmark.  This value is
small compared with the band-gap width in
Fig.~\ref{fig:SI_Er_DLG_honeycomb_bandstructure}, allowing the calculation
to resolve the qualitative difference between in-gap and band-edge
propagation.  Larger damping smooths the contrast between these regimes but
does not change the central band-gap criterion: inside the gap there is no
on-shell propagating graphene-plasmon Bloch mode.

The calculation is performed for source and observation points at the
midplane between the two graphene sheets,
\(
x_z=x_z'=0.
\)
At this symmetry plane, the symmetric and antisymmetric plasmon branches have different
field parity.  For the symmetric branch, the scalar potential is even in
\(z\), so the midplane potential and in-plane electric field are finite.
This branch contributes naturally to an in-plane Green-function component,
which we denote \(G_{xx}^{\rm sym}\).  The corresponding symmetric-sector midplane form
factor for a plane-wave component is
\begin{equation}
P_+(q_\alpha)
=
\frac{
2e^{-q_\alpha W/2}
}{
1+e^{-q_\alpha W}
},
\label{eq:app_symmetric_midplane_factor}
\end{equation}
and the normalized in-plane field factor used in the numerical sum is
\begin{equation}
\mathcal F_{\alpha,n\bk,+}^{(x)}
=
-\I L Q_{\alpha x}
P_+(q_\alpha)
\phi_{\alpha}^{(n\bk,+)} .
\label{eq:app_symmetric_field_factor}
\end{equation}
The factor \(L\) is included only to make the plotted Green functions
dimensionless; all Green-function curves are normalized.

For the antisymmetric branch, the scalar potential is odd in \(z\).  Therefore
the scalar potential itself vanishes at the midplane, but the vertical
electric field remains finite.  This sector contributes naturally to a
midplane \(G_{zz}^{\rm asym}\) component.  The normalized antisymmetric-sector vertical-field form
factor is
\begin{equation}
Z_-(q_\alpha)
=
(q_\alpha L)
\frac{
2e^{-q_\alpha W/2}
}{
1-e^{-q_\alpha W}
},
\label{eq:app_antisymmetric_midplane_factor}
\end{equation}
with the limiting value
\[
Z_-(q_\alpha\rightarrow0)
=
\frac{2L}{W}.
\]
The antisymmetric field factor used in the mode sum is
\begin{equation}
\mathcal F_{\alpha,n\bk,-}^{(z)}
=
Z_-(q_\alpha)
\phi_{\alpha}^{(n\bk,-)} .
\label{eq:app_antisymmetric_field_factor}
\end{equation}
Equations~\eqref{eq:app_symmetric_field_factor} and
\eqref{eq:app_antisymmetric_field_factor} convert the scalar-potential Bloch
functions obtained from the Gon\c{c}alves-Peres plane-wave method into the
midplane electric-field amplitudes that enter the Green tensor.

For a source at \(\brho_D\) and an observation point
\(\brho_A=\brho_D+x_\rho\hat{\mathbf e}\), the modal field amplitude at
position \(\brho\) is
\begin{equation}
\mathcal F_{n\bk,\lambda}^{(\eta)}(\brho)
=
\sum_\alpha
\mathcal F_{\alpha,n\bk,\lambda}^{(\eta)}
e^{\I\bQ_\alpha\cdot\brho}.
\label{eq:app_mode_field_real_space}
\end{equation}
The numerical Brillouin-zone integral in
Eq.~\eqref{eq:app_green_mode_sum_reduced} is performed by a midpoint grid
over the reciprocal primitive cell.  The plotted distance coordinate is the
physical lateral separation \(x_\rho\), not a lattice-index distance.

Because the absolute electromagnetic mode normalization is not included in
this figure-level plasmonic-crystal calculation, we plot normalized Green
functions.  The envelope used below is
\begin{equation}
|G_R^{\rm pl}|
=
\left[
|G_{xx}^{\rm sym}|^2
+
|G_{zz}^{\rm asym}|^2
\right]^{1/2},
\label{eq:app_green_envelope_reduced}
\end{equation}
and the complex response shown for phase information is the illustrative
equal-weight sum
\begin{equation}
G_{\rm eq}=G_{xx}^{\rm sym}+G_{zz}^{\rm asym}.
\label{eq:app_green_equal_weight}
\end{equation}
A common normalization is used for the three probe frequencies so that the
relative enhancement near the band edges and the suppression inside the
plasmonic band gap are visible.

\subsection{Directional cuts through the Green function}
\label{app:directional_green_cuts}

We evaluate the midplane Green function along three crystallographic
directions of the honeycomb superlattice:
\begin{equation}
\brho_A-\brho_D
=
x_\rho\hat{\mathbf e}_{a_1},
\qquad
x_\rho\hat{\mathbf e}_{a_2},
\qquad
x_\rho\hat{\mathbf e}_{a_1+a_2}.
\label{eq:app_green_directions}
\end{equation}
The direction vectors are normalized internally.  Thus the
\(a_1+a_2\) curve is a directional cut along the diagonal lattice
direction, while the horizontal axis remains the physical separation
\(x_\rho\) in nm.

The three probe frequencies are chosen from the plasmonic band structure in
Fig.~\ref{fig:SI_Er_DLG_honeycomb_bandstructure}:
\[
\omega=\omega_v,
\qquad
\omega=\omega_{\rm Er},
\qquad
\omega=\omega_c,
\]
where \(\omega_v\) is the top of the lower plasmon band and \(\omega_c\)
is the bottom of the upper plasmon band.  The Er frequency lies inside the
gap,
\begin{equation}
\omega_v<\omega_{\rm Er}<\omega_c .
\label{eq:app_Er_gap_condition_reduced}
\end{equation}
The comparison between \(\omega_v\), \(\omega_{\rm Er}\), and
\(\omega_c\) tests whether the in-gap Green function is genuinely
evanescent, rather than simply damped by finite \(\gamma_{\rm pl}\).

\subsection{In-gap versus band-edge Green-function decay}
\label{app:green_decay_results_reduced}

Figures~\ref{fig:app_green_envelope_valence_top},
\ref{fig:app_green_envelope_Er_transition}, and
\ref{fig:app_green_envelope_conduction_bottom} show the normalized
directional Green-function envelopes at three probe frequencies: the top
of the lower plasmon band, the Er transition frequency inside the gap, and
the bottom of the upper plasmon band.  The corresponding complex Green
functions are shown separately in
Figs.~\ref{fig:app_green_complex_valence_top},
\ref{fig:app_green_complex_Er_transition}, and
\ref{fig:app_green_complex_conduction_bottom}.  The envelope figures are
shown in single-column format because they contain one compact comparison
plot.  The complex-response figures are shown in two-column format because
they contain three directional panels.

At the Er transition, the Green function decays approximately
exponentially with \(x_\rho\), because there is no real-\(\bk\) plasmon
mode at that frequency.  Equivalently, the response is controlled by
complex-\(\bk\) evanescent Bloch waves.  At the top of the lower plasmon
band and at the bottom of the upper plasmon band, real or nearly real Bloch
plasmons become available.  The Green function then becomes more extended
and can show slower, oscillatory, or band-edge-enhanced decay.

The comparison demonstrates the central mechanism.  The same metagate
structure that opens the plasmonic band gap suppresses the on-shell
plasmonic loss channel at \(\omega_{\rm Er}\).  The Er-Er interaction is
therefore mediated by the reactive, off-shell part of the retarded Green
tensor rather than by a dissipative propagating plasmon.  This is the
plasmonic-crystal analogue of placing the transition frequency inside a
photonic band gap: propagating modes are removed at the transition
frequency, while virtual modes still contribute to the retarded Green
tensor.

\bibliographystyle{apsrev4-2}
\bibliography{universal_FRET_refs_with_SI}

\end{document}